\def\year{2021}\relax
\documentclass[letterpaper]{article} 
\usepackage{aaai21}  
\usepackage{times}  
\usepackage{helvet} 
\usepackage{courier}  
\usepackage[hyphens]{url}  
\usepackage{graphicx} 
\usepackage{csquotes}
\usepackage{amssymb}
\usepackage{rotating}
\usepackage{amsmath}
\usepackage{natbib}  
\usepackage{caption} 
\usepackage{array}
\usepackage{setspace} 
\usepackage{makecell}
\usepackage[flushleft]{threeparttable}
\usepackage{booktabs,caption}
\usepackage[toc,page]{appendix}
\usepackage{subcaption}
\usepackage{rotating}
\usepackage{multirow}
\usepackage[T1]{fontenc}

\usepackage[usenames,dvipsnames]{color}

\title{A Fine-Grained Analysis of Public Opinion toward Chinese Technology Companies on Reddit}
\author {
    Enting Zhou\textsuperscript{*}, Yurong Liu\textsuperscript{*}, Hanjia Lyu, Jiebo Luo\\
}
\affiliations {
    University of Rochester\\
    \{ezhou12,yliu217\}@u.rochester.edu, hlyu5@ur.rochester.edu, jluo@cs.rochester.edu\\
    \textsuperscript{*} these authors contributed equally
}

\begin{document}

\maketitle

\begin{abstract}
In the face of the growing global influence and prevalence of Chinese technology companies, governments worldwide have expressed concern and mistrust toward these companies. There is a scarcity of research that specifically examines the widespread public response to this phenomenon on a large scale. This study aims to fill in the gap in understanding public opinion toward Chinese technology companies using Reddit data, a popular news-oriented social media platform. We employ the state-of-the-art transformer model to build a reliable sentiment classifier. We then use LDA to extract the topics associated with positive and negative comments. We also conduct content analysis by studying the changes in the semantic meaning of the companies' names over time. Our main findings include the following: 1) Notable difference exists in the proportions of positive comments (8.42\%) and negative comments (14.12\%); 2) Positive comments are mostly associated with the companies’ consumer products, such as smartphones, laptops, and wearable electronics. Negative comments have a more diverse topic distribution (notable topics include criticism toward the platform, dissatisfaction with the companies’ smartphone products, companies’ ties to the Chinese government, data security concerns, 5G construction, and general political discussions); and 3) Characterization of each technology company is usually centered around a particular predominant theme related to the company, while real-world political events may trigger drastic changes in users' characterization.

\end{abstract}

\section{Introduction}

In the past few years, China has been quickly building up its global influence, especially in the realm of technology. The internationalization of Chinese technology companies has also contributed to the growing global influence. The ubiquitous presence of Chinese technology companies has unnerved governments worldwide, especially in the West. Many governments have put forward legislation to tightly regulate Chinese technology companies’ operations in their countries. In some extreme cases, governments have explicitly forbidden certain Chinese technology companies from providing service in their countries.\footnote{\url{https://www.euractiv.com/section/digital/news/eu-countries-keep-different-approaches-to-huawei-on-5g-rollout/} [Accessed Jan. 8, 2022]} \footnote{\url{https://www.latimes.com/business/la-na-pol-trump-huawei-ban-20190515-story.html} [Accessed Jan. 8, 2022]}  

Previous studies investigated  governments’ reactions to Chinese technology companies and sought to explain the governments’ mistrust toward these companies. \citet{huawei_mwz} argued that the distrust and restriction imposed on Chinese technology companies stem from the fact that technology has increasingly strategic importance in today’s world, especially in the realms of national defense, commerce, and social organization. \citet{hue_cry_huawei} found that the mistrust toward Chinese technology firms is not only ideological but also commercial. The deliberate mistrust is built in efforts to suppress China's growing economic presence in the global market with technology firms, such as Huawei, as the face of such rising economic power. 

Few previous studies have focused on the large-scale public reaction to the growing influence and presence of Chinese technology companies. Therefore in this study, we attempt to study public opinion toward Chinese technology companies, especially ones with great influence outside China, such as Huawei, Tencent, ByteDance (Tiktok), and Alibaba. This topic is itself very sensitive to daily real-world events. We choose Reddit as our source of data. It is one of the largest news-oriented social media platforms, serving as a good data source for capturing and analyzing public opinion on such a news-sensitive topic. Reddit has 530 million monthly active users, and a diverse user base across the world, providing us with a copious amount of data.\footnote{\url{https://backlinko.com/reddit-users} [Accessed Jan. 8, 2022]} Reddit also has a more flexible comment length constraint - up to 40,000 characters per post compared to merely 280 characters on Twitter. As a result, Reddit stands out as a superior platform for retrieving high-quality conversational data and discussions. Another special advantage of Reddit is that unlike Twitter or other major social media platforms, Reddit does not require users to fill in their personal information, therefore creating an anonymous environment on Reddit, which is crucial in their effort to safeguard users' freedom of speech~\cite{reddit_freespeech}. Reddit has been adopted by many researchers as a reliable resource for large-scale, relatively high-quality posts on a topic of broad interest (e.g., COVID vaccine, rape, e-cigarette)~\cite{lyu_reddit, kamarudin2018study, chen2020social}. In this paper, we collect data from Reddit to examine public sentiment in the discussions of Chinese technology companies. We aim to answer the following research questions:
\begin{itemize}
    \item \textbf{RQ1}: What sentiment do Reddit users generally express in discussions of Chinese technology companies? 
    \item \textbf{RQ2}: What are the topics associated with the positive and negative sentiments?
    \item \textbf{RQ3}: What are the dominant features in the discussions of a particular company? Do they change over time?
\end{itemize}

We approach these questions using a series of computational methods. We adopt a human-guided machine learning framework based on a transformer model to classify comments’ sentiments. We model the latent topics of both positive and negative comments. We investigate the users’ characterization of each particular company by training a set of word2vec models to generate word embeddings and calculating the most semantically similar words to the company names. To summarize, we find:
\begin{itemize}
    \item The public often show more negative attitude than positive attitude toward Chinese technology companies.
    \item Positive comments are mainly about companies' products, while the topics are more diverse in the negative comments.
    \item The results of the word embeddings show predominant themes in the discussions regarding Chinese technology companies. In addition, we have identified various real-world events that have a significant impact on how the general public perceives these companies.
\end{itemize}
Compared to previous studies focusing on the relations between Chinese technology companies and the Western governments~\cite{huawei_mwz,hue_cry_huawei}, we contribute to a better understanding of \textbf{large-scale public sentiment} toward Chinese technology companies. To our best knowledge, this is the first study that investigates public opinion toward Chinese technology companies. Additionally, we make our dataset publicly available to the research community to facilitate future work.

\section{Related Work}
Traditional methods of analyzing public opinion rely on soliciting responses to a poll, which have several disadvantages. They are expensive and time-consuming. Their quality may easily suffer from poor sampling and a low response rate~\cite{polling}. With growing influence, Social media platforms have become an alternate source for researchers to obtain massive text-based datasets. They provide easy access to public opinion in real time~\cite{rating}, and have enabled researchers to shed new light on human behavior and opinion related to major issues of global importance~\cite{polling}. There has been a significant amount of research in public opinion that analyzed social media data. For instance, \citet{climate} used Twitter data to study public sentiment toward the climate change problem. They have uncovered that on Twitter, the responses to climate change news are predominately from climate change activists rather than climate change deniers, indicating that Twitter is a valuable resource for the spread of climate change awareness. More recently, \citet{vaccine} focused on public opinion on vaccines in the face of the COVID-19 pandemic. Empowered by the rich information provided on social media platforms, they found correlations between user demographics and their attitudes toward vaccine uptake.

Many studies have been conducted to monitor and analyze the opinion on Reddit. Some studies collected posts under a small number of selected related subreddits. For instance, \citet{shen} investigated anxiety on Reddit by comparing the posts under a group of anxiety-related subreddits to the posts under a control group of subreddits that are unrelated to anxiety.  \citet{Farrell} analyzed the phenomenon of misogyny by collecting posts under a set of carefully selected subreddits around the topics of men’s rights and difficulty in relationships. Other studies have used off-the-shelf datasets that contain all the posts on Reddit in a given period. For instance, \citet{Soliman} studied the characterization of the political community using a readily available dataset that includes all submissions and comments on Reddit from 2005 to 2018. Such a data collection method provides the scale that we require for our study, but it introduces enormous difficulty to distinguish between relevant and irrelevant content to our study subjects. \citet{lyu_reddit} studied public sentiment toward COVID-19 vaccines. In this paper, we adopt their methodology to perform keyword research on a sitewide basis, which provides a breadth of data while keeping data noise at a minimum.

Computational methods have been employed to study public sentiment using social media data. To study public sentiment toward COVID-19 vaccines, \citet{vaccine} applied the state-of-the-art transformer model - XLNet to mining opinions. \citet{goeatbat} aimed at revealing online Sinophobic behaviors during the COVID-19 pandemic using word embeddings. By modeling Chinese-related terms on Twitter and 4chan forums as word vectors, they found that Chinese-related terms are associated with racial slurs on both Twitter and 4chan, thus revealing the rise of Sinophobic behaviors in a cross-platform manner. By analyzing the word embeddings’ temporal change, they discovered new racial slurs related to China and the tendency to blame China and Chinese people as the pandemic escalates. In our study, we apply a similar method to understanding public opinion about Chinese technology companies.

\section{Datasets}

A common problem faced by all approaches that attempt to measure public opinion is the problem of selecting appropriate samples that will generalize to the public population. Inappropriate selection of the sample will expose the study to self-selection bias that limits the generalizability outside the study sample. Social media study might be more prone to this problem, as people’s motivation to participate in social media conversations regarding a particular topic may be diverse and hard to be distinguished as opposed to conventional survey research, which is able to assert study subjects’ motivation directly~\cite{10.1093/poq/nfab021}. Furthermore, Reddit itself underwent significant political polarization since the 2016 US election, resulting not from increasing individual polarization but from system-level shifts driven by the arrival of new users~\cite{10.1145/2818048.2820078}. Therefore, while we acknowledge that the study population of this study may have polarization issues and inherent bias, we think likewise as \citet{10.1093/poq/nfab021} that this built-in bias in the study population will facilitate us to identify and better understand issues that may be previously unknown, unintentionally ignored or marginalized in conventional studies. 

\subsection{Data Collection}

Reddit consists of individual micro-communities that are based on certain topics (i.e., r/worldnews, r/technology), which are referred to as subreddits. Under each subreddit, users can make posts, known as submissions in Reddit. Users can also make comments under submissions. To obtain the data for our study, we collect publicly available content from Reddit using a list of keywords, which contain the names of Chinese technology companies and the names of chief executive officers (CEOs). Note that we choose not to collect submissions from Reddit, because compared to comments, most submissions are opinion-neutral content, such as news and questions, and usually do not indicate users’ opinions toward our study subjects. Therefore, we choose {\it users’ comments} as our data source.

To collect users’ comments, we employ a Reddit API Wrapper called PRAW.\footnote{\url{https://praw.readthedocs.io/en/latest/} [Accessed Jan. 8, 2022]} We perform keyword searches with names of the companies and the names of their CEOs (e.g., “Tencent”, “Huawei”, “Ma Huateng”, “Ren Zhengfei”) on a sitewide basis. We pull comments over two years, from November 1, 2019 to November 1, 2021. Duplicate, non-English, and automatic moderator comments are pruned from the dataset. Our final dataset contains 294,610 comments from 172,453 distinct authors.

\subsection{Data Preprocessing}
To prepare the data for further sentiment classification and language modeling, we perform a text cleaning process. We convert all words to lowercase and remove all uniform resource locators (URLs) and numbers from the text. We use a dictionary of English stop words provided by the Natural Language Toolkit (NLTK)\footnote{\url{https://www.nltk.org/} [Accessed Jan. 8, 2022]} to remove all stop words from the text. Additionally, we perform a text lemmatization using NLTK.

\begin{table*}[t]
\centering
\caption{The labeling scheme.}
\begin{tabular}{|c|l|}
\hline
Class & \multicolumn{1}{|c|}{Description} \\
\hline
Positive & \begin{tabular}[c]{@{}l@{}}i. Expressing positive sentiment toward the Chinese tech companies themselves\\ ii. Expressing positive sentiments toward the tech companies’ products (i.e., social platforms, \\smartphones, laptops, \textit{etc}.)\end{tabular} \\ \hline
Negative & \begin{tabular}[c]{@{}l@{}}i. Expressing negative sentiment toward the Chinese tech companies themselves\\ ii. Expressing negative sentiment toward Chinese tech companies’ products (i.e., social platforms, \\smartphones, laptops, \textit{etc}.)\\ iii. Promoting/arguing in favor of conspiracy theories about the Chinese tech companies\\ iv. Advocating for a ban or sanction toward the Chinese tech companies\end{tabular} \\ \hline
Neutral/Irrelevant & \begin{tabular}[c]{@{}l@{}}i. News regarding Chinese tech companies with no written opinion from the commenters\\ ii. Including Chinese tech companies and the commenters’ opinions, but the focus is \\something else (i.e., politics, economics, \textit{etc}.)\\ iii. Comments with {\it neutral} opinion toward Chinese technology companies \\ iv. Comments/questions on Chinese tech companies entities but with unclear meanings\end{tabular} \\ \hline
\end{tabular}

\label{tab:table1}
\end{table*}

\section{Text Sentiment Classification}

\subsection{Method}

XLNet~\cite{xlnet} is a generalized autoregressive pretraining method that can capture left and right contexts jointly in sentences. It was claimed that XLNet outperforms BERT (Bidirectional Encoder Representations from Transformers)~\cite{bert}, a pretrained language model with transformer architecture that is designed to perform downstream NLP tasks after fine-tuned, on 20 tasks such as question answering, natural language inference, sentiment analysis, and document ranking. Previous studies have used XLNet models fine-tuned for emotion classification, sentiment analysis~\cite{xlnetOP}, and classification of censored tweets~\cite{xlnetCensorCn}, \textit{etc}. We employ XLNet Base for our task and limit each \textit{comment\_body} to the first 512 tokens. Next, we use the Adam optimizer to fine-tune our XLNet model for three epochs. It predicts a probability for each of the three possible categories (i.e., positive, negative, neutral/irrelevant) for \textit{comment\_body}. We have also experimented with VADER~\cite{vader}, a lexicon and rule-based sentiment analysis tool which is specifically designed for social media posts.

\subsection{Data Labeling}
To study the sentiment of each comment toward a Chinese technology company, we classify each comment into three categories: (1) positive, (2) negative, and (3) neutral/irrelevant according to the labeling scheme in Table~\ref{tab:table1}. Our initial collected data presents a challenging class imbalance problem among these three classes where the majority of the data is irrelevant to the opinion toward these companies. To address this problem, we adopt the approach by \citet{vaccine} to employ a human-guided machine learning framework based on the state-of-the-art transformer model.

More specifically, we build our initial training dataset by randomly sampling 1,600 comments from the entire corpus of 294,610 comments. For each comment, two researchers independently read and label the comment with one of the three categories presented in Table~\ref{tab:table1}. If the two labels given by the two researchers are different, then a third researcher would discuss with the group to determine the consensus label of the comment. It is worth noting that three researchers' decisions reach a Fleiss’ Kappa score of 0.77, which indicates a good agreement among the annotated labels.  

\begin{figure}[t]
\centering
\includegraphics[width=0.7\columnwidth]{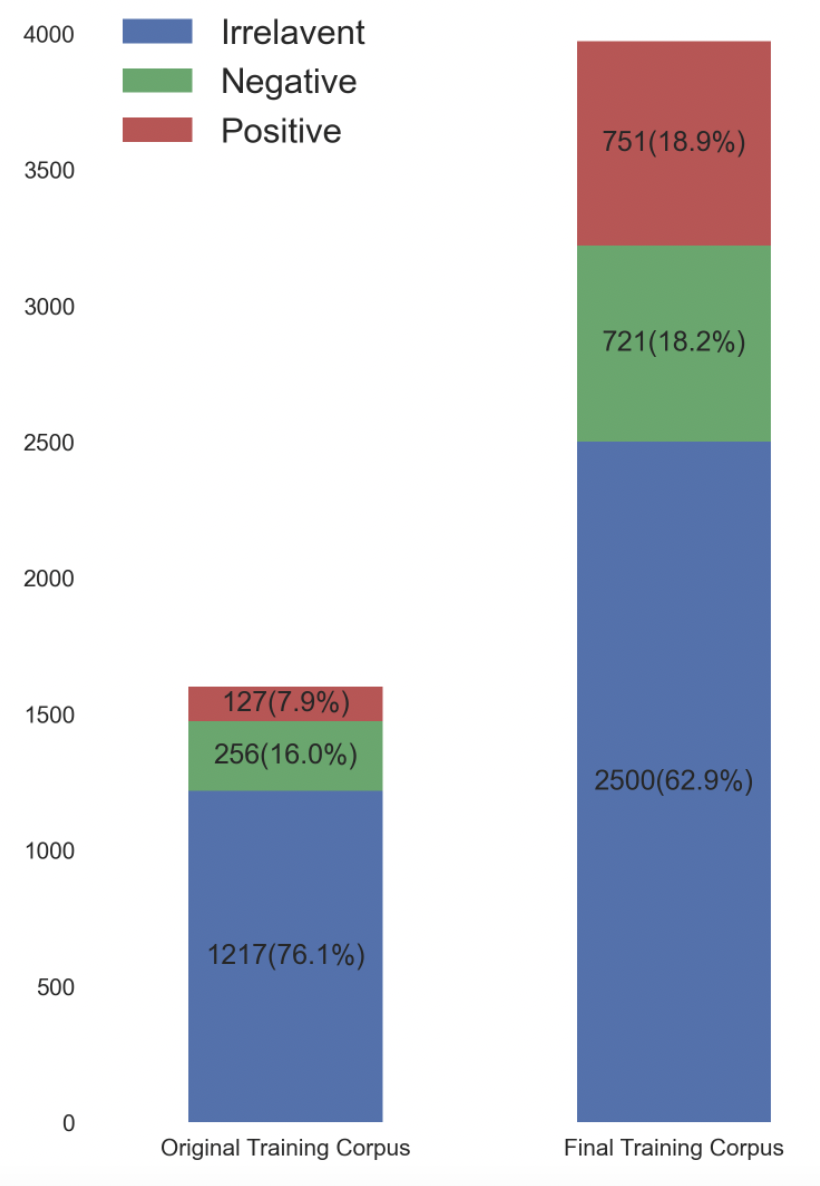}
\caption{Distributions of three categories of the initial training corpus and the final training corpus.}
\label{tab:figure1}
\end{figure}

We train an XLNet model $H$ using these 1,600 labeled comments as the initial training corpus ($n=1600$). However, due to the severe imbalance issue, $H$ performs poorly at predicting positive comments. We then use $H$ to construct a new batch of training corpus of 1,000 comments. Out of these 1,000 comments, 45\% are the comments that $H$ predicts to be most likely positive, 45\% are the comments that $H$ predicts to be most likely negative, and 10\% are randomly sampled to increase diversity. The new batch of 1,000 comments is labeled by researchers and added to our original training corpus. Next, we train $H$ using this new training corpus ($n=2600$). This entire process is considered one iteration. We repeat two iterations before obtaining our final training corpus, with balanced data between positive and negative categories. The distributions of three categories of the initial and final training corpus are displayed in Fig.~\ref{tab:figure1}. This framework actively searches for the most possible negative and positive comments to increase the size of our training data and strike toward a more balanced class distribution.

\subsection{Evaluation}

\begin{table}[t]
\centering
\caption{Performance of the XLNet model and VADER.}
\begin{tabular}{|wc{0.8cm}|wc{2.5cm}|wc{1cm}|wc{0.9cm}|wc{1cm}|}
\hline
Model & Class & Precision & Recall & F1-score \\
\hline
VADER & \begin{tabular}[c]{@{}l@{}}Irrelevant \\ Negative\\ Positive\\ Overall (Weighted)\end{tabular} & \begin{tabular}[c]{@{}l@{}}0.83\\ 0.36\\ 0.29\\ 0.65\end{tabular} & \begin{tabular}[c]{@{}l@{}}0.17\\ 0.59\\ 0.90\\ 0.38\end{tabular} & \begin{tabular}[c]{@{}l@{}}0.28\\ 0.45\\ 0.44\\ 0.34\end{tabular} \\
\hline
XLNet & \begin{tabular}[c]{@{}l@{}}Irrelevant\\ Negative\\ Positive\\ Overall (Weighted)\end{tabular} & \begin{tabular}[c]{@{}l@{}}0.86\\ 0.62\\ 0.74\\ 0.79\end{tabular} & \begin{tabular}[c]{@{}l@{}}0.80\\ 0.69\\ 0.82\\ 0.78\end{tabular} & \begin{tabular}[c]{@{}l@{}}0.83\\ 0.65\\ 0.78\\ 0.78\end{tabular} \\ 
\hline
\end{tabular}
\label{tab:table2}
\end{table}

Table~\ref{tab:table2} summarizes our final model's performance. We use the ``weighted'' F1-score to evaluate the overall performance. An overall F1-score of 0.78 is obtained by the XLNet model, as well as similar F1-scores for positive and negative classes, which are sufficiently reliable for our further analysis. The XLNet model outperforms the lexicon and rule-based sentiment analysis tool VADER, which may be because (1) the pre-built sentiment analysis tool cannot handle the complex semantic meanings, and (2) sentiment is not opinion (i.e., a comment can contain many negative words but express positive opinion toward the Chinese technology companies). This further supports the necessity of our human-guided machine learning framework.

\begin{figure} [h]
\includegraphics[width=0.91\columnwidth]{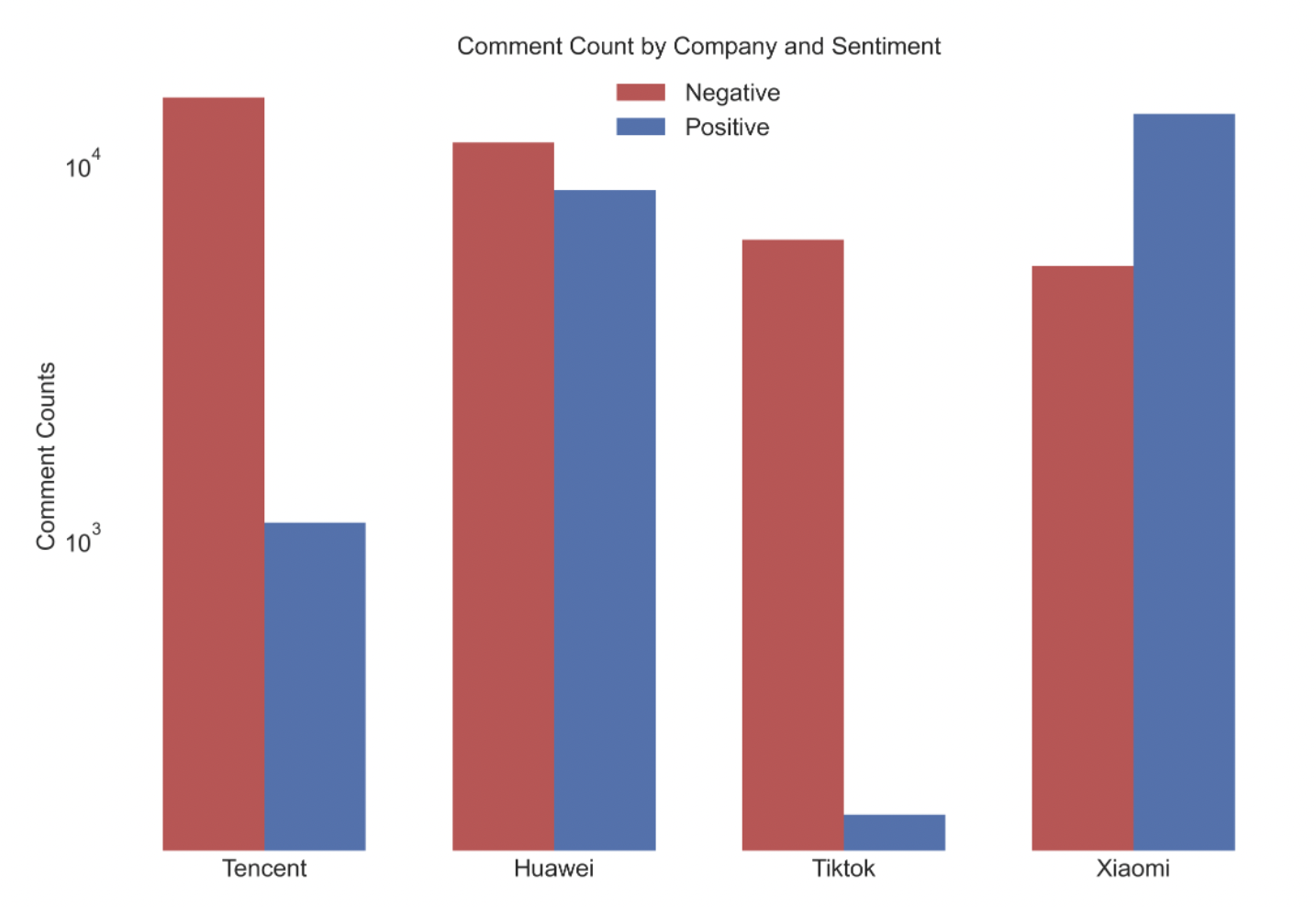}
\caption{Comment counts in logarithmic scale by sentiment.}
\label{tab:figure3}
\end{figure}

Based on the classification results by our fine-tuned XLNet model on the entire dataset, we plot a side-by-side bar chart (Fig.~\ref{tab:figure3}) to compare the difference between the number of positive comments and the number of negative comments in the overall discussion and specific discussion on the top four most mentioned technology companies. We find, while our study samples comprise primarily (77\%) of neutral or irrelevant content, there are notable differences in the proportions of positive comments (8.42\%) and negative comments (14.12\%), where there are 68\% more negative comments than positive comments. This finding gives insight into the general sentiment the Reddit community holds toward Chinese technology companies. Our subsequent topic analysis and content analysis will delve into the underlying reasons for these variations in sentiment, with meticulous attention to detail. One interesting observation can be made from the company-wise comparison: Two companies that have the most positive comments, Xiaomi and Huawei, are all major producers of consumer electronics such as smartphones and laptops, while the two companies that have predominantly negative comments, Tencent and Tiktok, are companies that offer online services, such as social media platforms and video games.

\section{Topic Modeling}
\begin{figure*} [t]
\centering
\includegraphics[width=\linewidth]{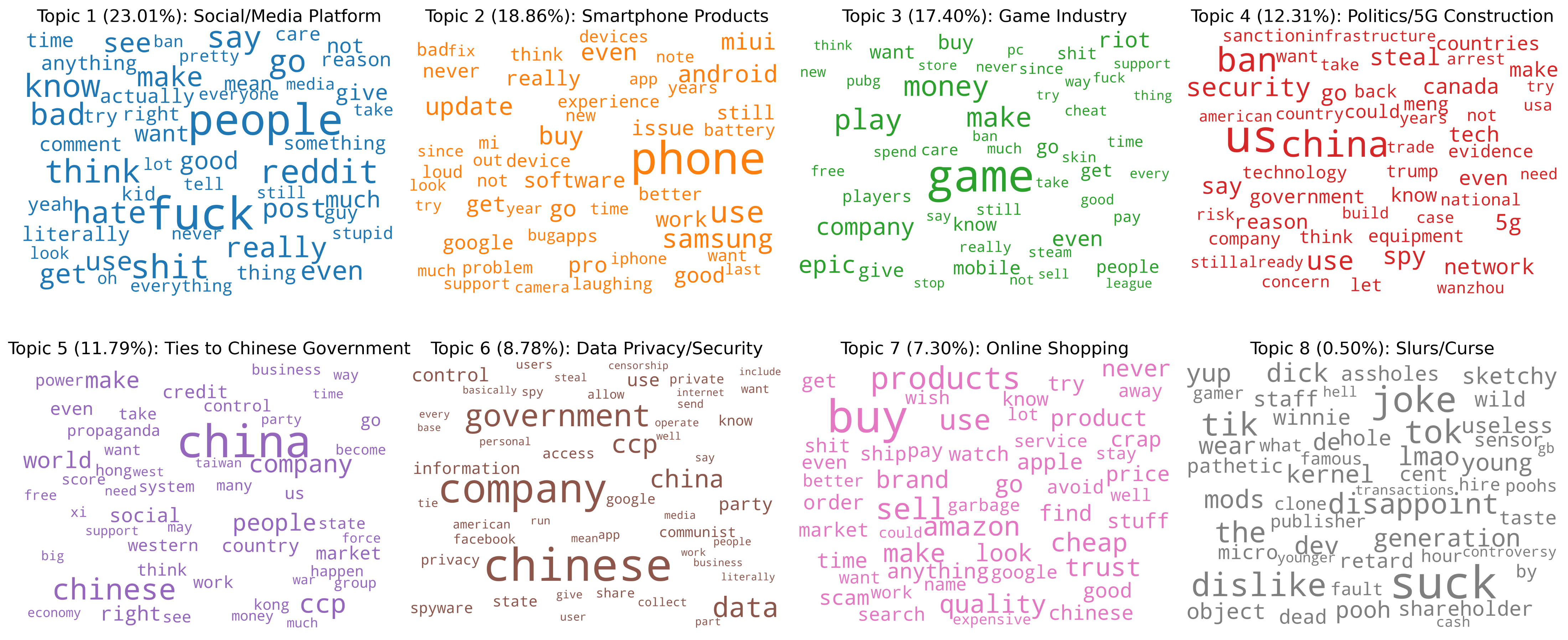}
\caption{Word cloud of the negative comment topics.}
\label{tab:figure4}
\end{figure*} 

\begin{table*}[h]
\centering
\caption{Eight topics of negative Reddits generated by LDA and their associated example comments.}
\scalebox{1.0}{
\begin{tabular}{|l|l|l|l|}
\hline
ID & Topics                                     & \% of comments & Example Comment  \\
\hline
1 & Social/Media Platform         & 23.01\%        & \begin{tabular}[c]{@{}l@{}}“i am tired of tiktok with their insensitive trends. \\ the community is terrible, i see a ton of racist,  \\ homophobic and abliest people on that app. i dont\\  mind tiktok but i hate the community, it reminds \\ me of school becos of how similar they are.”\end{tabular}                                                                                   \\
\hline
2 & Smartphone Products & 18.86\%        & \begin{tabular}[c]{@{}l@{}}“I would NEVER ever buy huawei now. I have \\ Mate series phone, although I get upgrades still. \\ One of the upgrade has taken by ability to change\\  my launcher from default to anything I want...."\end{tabular}                                                                                                                                              \\
\hline
3 & Gaming Industry       & 17.40\%        & \begin{tabular}[c]{@{}l@{}}“Tencent really only cares about the skins and \\ them making money and trying to make people \\ spend money with this blackpink whatever this is \\ all they really need to do is start banning this \\ cheaters the devs need to know if the game has \\ like no hackers to a little people would actually \\ consider spending money in the game.”\end{tabular} \\
\hline
4 & Politics/5G construction                   & 12.31\%        & \begin{tabular}[c]{@{}l@{}}“China will just steal the ip of those exports once\\  they can and produce it themselves huawei is \\ based off nortel a canadian company that fell \\ because of industrial espionage they didnt even \\ change the stolen source code. China forcing nations\\  on 5g just goes to prove it will be used for spying”\end{tabular}                               \\
\hline
5 & Ties to Chinese Government                 & 11.79\%        & \begin{tabular}[c]{@{}l@{}}“Chinese propaganda goes a long way. all this \\ acting to make us believe all is well and people are \\ happy instead of showing the massacre of Uighurs. \\ xi jinping and tencent can go to hell...”\end{tabular}               \\
\hline
6 & Data Privacy/Security                      & 8.78\%         & \begin{tabular}[c]{@{}l@{}}“Every major company in China is required to have\\  a party board member and every Chinese company\\ by law in China has to share it's data with the \\ government. Also Tencent has significant ties to CCP\\  including having built part of the Chinese government\\ state surveillance platform.”\end{tabular}                                                \\
\hline
7 & Online Shopping                            & 7.30\%         & \begin{tabular}[c]{@{}l@{}}“Me too. I have used Alibaba and wish, all about the \\ same wait times unless you pay gobs. The shitty part\\ is that the refund process can take way longer."\end{tabular}                                                                                                                                                              \\
\hline
8 & Slur/Curse                                 & 0.50\%         & “Yup. Absolutely. F**k TikTok.” \\
\hline
\end{tabular}
}
\label{tab:table_n}
\end{table*}

\subsection{LDA} Latent Dirichlet allocation (LDA) by \citet{lda} is a topic modeling method that has been used extensively in studies on social networks and microblogging environments~\cite{lda_1}. \citet{lda_2} found that LDA can achieve better performance in a short text context compared to other topic modeling methods, such as Latent Semantic Analysis (LSA)~\cite{LSA}, Principal Component Analysis (PCA)~\cite{PCA}, and random projection (RP)~\cite{RP}. It also produces higher quality topics and more coherent topics than other topic modeling methods. Therefore, we choose LDA as our method for topic modeling. 

To characterize different topics associated with positive and negative comments, we apply LDA separately for negative and positive comments. To better differentiate possible topics, we remove the names of the companies we are studying from the processed text corpus. To determine the optimal number of topics, we train a set of LDA models, compute their coherence scores, and read the representative keywords. For negative comments, the number of topics is set to 8. The model has a coherence score of 0.54. The number of topics for positive comments is set to 2 which gives rise to a coherence score of 0.56.

\subsection{Result} 

Table~\ref{tab:table_n} itemizes the eight topics that are associated with negative sentiment and their example comments. Fig.~\ref{tab:figure4} visualizes the keywords in each negative comment topic in the form of a word cloud and displays the percentage of comments belonging to each topic. We adopt a common approach in topic modeling by manually assigning topic labels according to keywords in the topics generated from the model. We also categorize each comment to its dominant topic.

Topic 1 indicates that the most frequent topic, accounting for 23\% of the overall negative comments, is criticism toward social media platforms. The most heavily criticized companies are Tiktok and Tencent which account for up to 80\% of the total comments on this topic. Many commenters express sharp negative sentiment toward these two companies’ communities and online platforms with swear words such as “hate”, “stupid”, “suck”, \textit{etc}. While TikTok's platform content is frequently negatively commented on as stupid and cringe, Tencent is blamed for its poor control of cheaters in its gaming communities where the anti-cheat system is not efficient enough. Another interesting observation is that ``Reddit'' appears as a keyword in the topic. We find that the Reddit community itself is commonly mentioned in this topic, and the relationships between Reddit and these two companies are very different. Since Tencent has invested in Reddit, certain Reddit users argue that the involvement of Tencent has introduced low-quality content and heavy censorship on the Reddit platform. On the other hand, Reddit users compare their platform with TikTok and imply that Reddit is a superior social media platform.

Topic 2 illustrates the second most frequent topic - criticism toward smartphone products of certain companies such as Huawei and Xiaomi. The third most frequent topic corresponds to negative sentiment toward the overarching influence of Chinese technology companies, Tencent in particular, on the gaming industry. A prevalent criticism on this front is that Tencent's monopoly and significant profits from investing in major gaming companies have resulted in little to no positive impact on the gaming industry, and in some cases, even a negative impact.

The next three topics focus on the negative sentiment toward the “Chinese” part of the companies. The fourth most frequent topic is the discussion about politics between China and foreign governments, the trade war with the US, and some recent events, such as Meng Wanzhou’s arrest in Canada.\footnote{\url{https://www.cnn.com/2018/12/05/tech/huawei-cfo-arrested-canada/index.html} [Accessed Jan. 12, 2022]} We can observe how these political events' influence may help shape the public’s view on Huawei from some example comments. For instance, a few comments express mistrust toward Huawei in response to two Canadian citizens being detained by the Chinese government. Another important subtopic is Chinese technology companies’ involvement in 5G construction. Huawei, as the leading company of the world’s telecommunication providers, becomes the target of these criticisms.
 
The fifth most frequent topic is the link between these companies to the Chinese government. Many representative comments express negative sentiments toward these companies because of their alleged close relationship with the Chinese government. Some comments hold these companies responsible for assisting the Chinese government’s alleged “mass surveillance” or “genocide of the Uighurs”. Topic 6 talks about how the comments perceived these companies as a risk or danger to users' data privacy. Many associated comments refer to these companies as the “state surveillance platform”, “national threat concerns” or “spyware created by the CCP (Chinese Communist Party)”.

Interestingly, we can observe two major groups of topics. The first, second,  third, and seventh topics could be grouped together, comprising 2/3 of all negative comments, because they are mostly concentrated on the “technology” aspect of these companies, while the fourth, fifth, and sixth topics could be grouped together, comprising 1/3 of all negative comments, because they are concentrated on the “Chinese part” of Chinese technology companies.   

\begin{figure*} [t]
\centering
\includegraphics[width=0.85\linewidth]{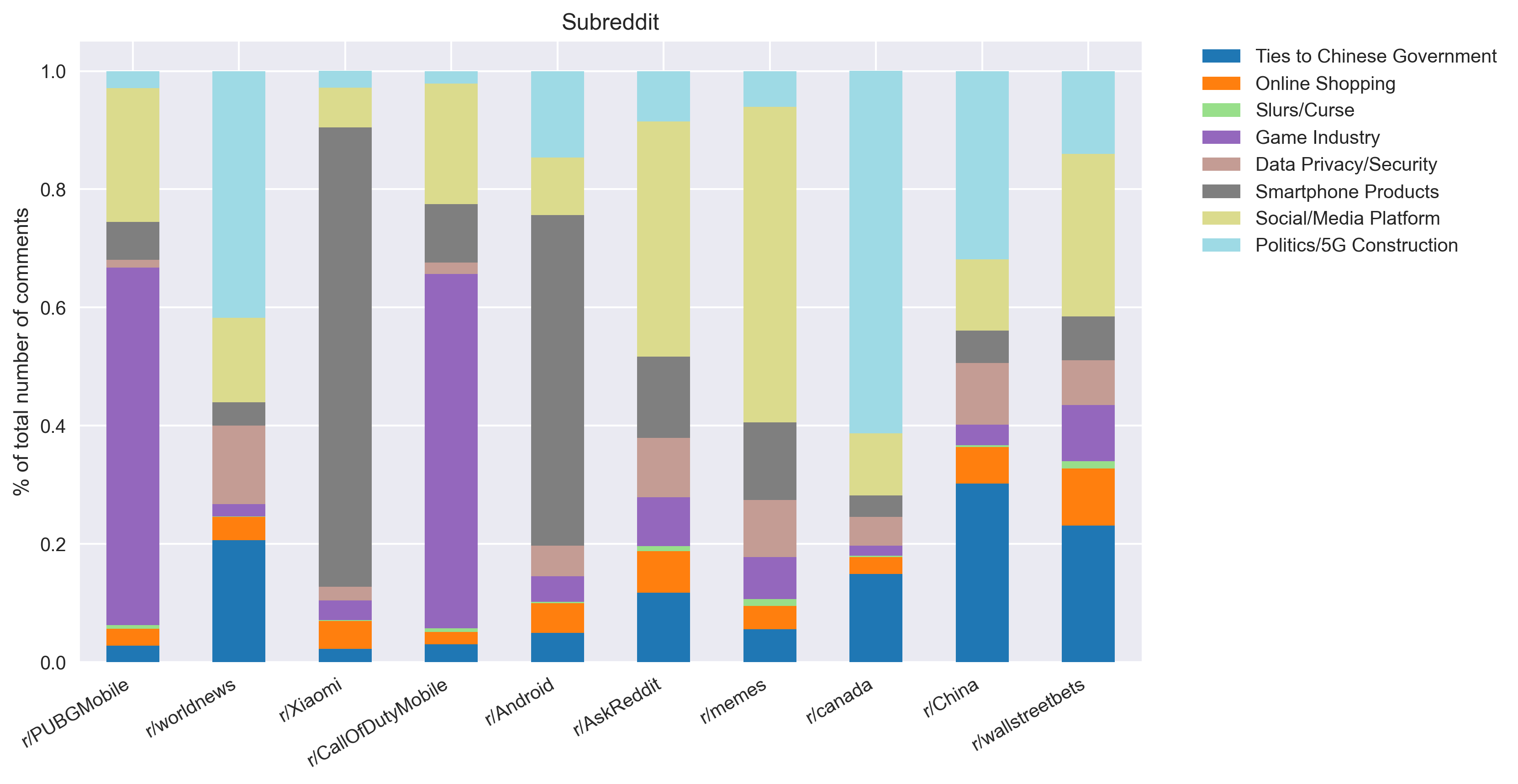}
\caption{Distributions of negative sentiment-associated topics across the top 10 subreddits.}
\label{tab:subreddit}
\end{figure*} 

We have also analyzed the topic distributions from a cross-sectional perspective. Fig.~\ref{tab:subreddit} shows the distributions of different topics by the top 10 subreddits that have the most negative comments. The topic distributions correspond to the subreddit’s theme well. Subreddits that are centered around games, such as r/PUBGMobile and r/CallOfDutyMobile, have the largest proportions of negative comments of the gaming industry topic, whereas subreddits that are centered around smartphones, such as r/Xiaomi and r/Android, have the largest proportions of negative comments of phone complaints and criticism topic. More importantly, subreddits that have a more political theme, such as r/worldnews, r/Canada, and r/China, have apparently more discussions surrounding Chinese technology companies’ ties to the Chinese government, global politics, and 5G constructions. The coherence between topic distributions and subreddits' themes shows the good performance and robustness of our topic model.

\begin{figure}[h]
\centering
\includegraphics[width=\columnwidth]{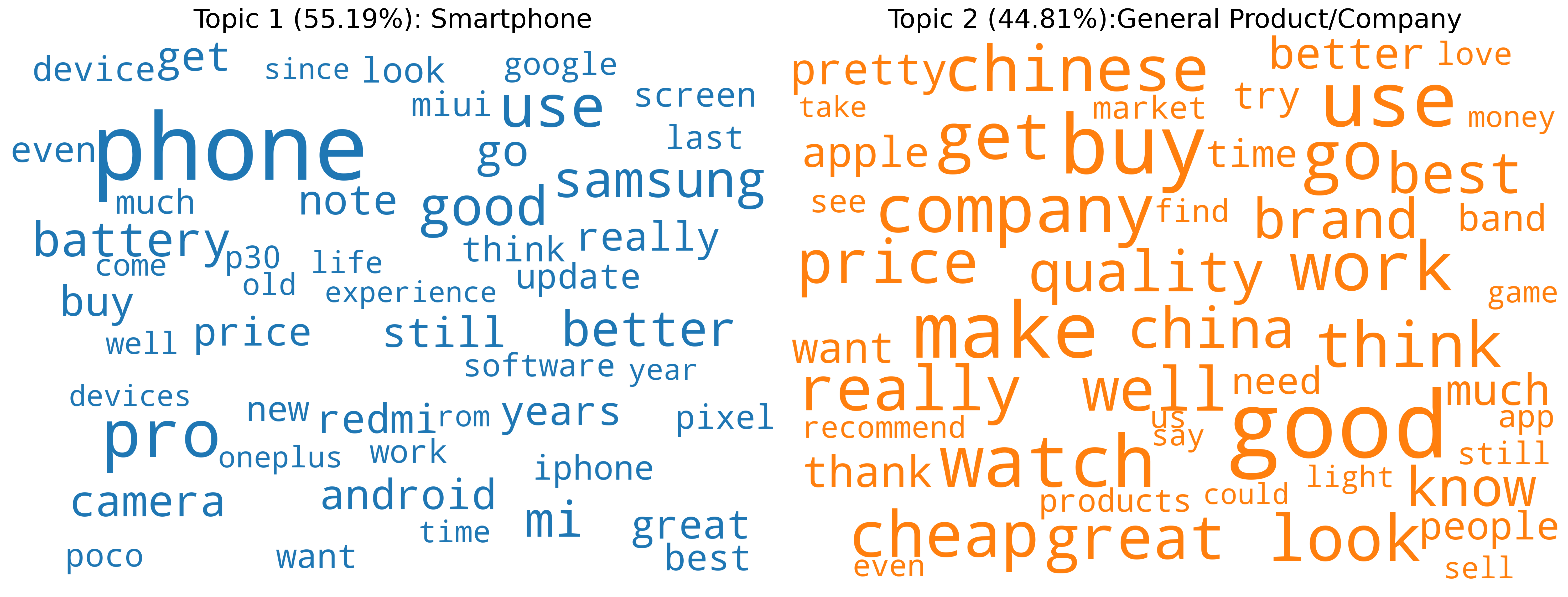}
\caption{Word cloud of the positive comment topics.}
\label{tab:figure5}
\end{figure}

Fig.~\ref{tab:figure5} shows the two topics that are associated with positive comments.
Compared to negative comments, positive comments are more monotonic. They are largely associated with the companies’ physical products. The positive comments are approximately equally divided between these two topics. The first one focuses on the general aspect of physical products, which constitutes $55.19\%$ of the total comments, and the second one is more specifically related to the companies’ smartphone products, which accounts for $44.81\%$. After further investigation, we find that the comments on both topics are mainly associated with Huawei and Xiaomi, the two consumer electronic manufacturers in the top four most mentioned Chinese technology companies. This also explains the large number of positive comments for Huawei and Xiaomi displayed in Fig.~\ref{tab:figure3}.  

\section{Temporal \& Content Analysis}

To further understand the context of the words and the changes of public opinion on Chinese technology companies over time, we employ the skip-gram algorithm~\cite{word2vec}. Specifically, we train a set of word2vec models on specific groups of comments, which reflect the most closely associated terms to each company during each month of our study period. For positive comments and negative comments, we train one word2vec model separately for each month. After training, we calculate the cosine similarity of the word vectors obtained from each model to study the similarity among words. We conduct an analysis of the top four mentioned companies, Huawei, Tencent, ByteDance, and Xiaomi. However, we discover the discourse around Xiaomi is mainly about its consumer electronics, which coincides with our finding in the previous section using LDA and offers no new insight, therefore we omit the discussion of Xiaomi in this section and focus solely on the other three companies.

\subsection{Case Study 1: Huawei}

\begin{table*}[htbp]
\centering
\scriptsize
\caption{Top 25 most semantically similar words to ``Huawei'' obtained from the monthly word2vec models of negative comments from November 1, 2019 to November 1, 2021. Words related to telecommunication technology, political entity, and network security are labeled red, and words related to consumer electronic products are labeled blue, respectively.}
\scalebox{.9}{
\begin{tabular}{|ll|llllllllll|}
\hline
\multicolumn{2}{|c|}{2019} & \multicolumn{10}{|c|}{2020} \\
\hline
\textbf{Nov} & \textbf{Dec} & \textbf{Jan} & \textbf{Feb} & \textbf{Mar} & \textbf{Apr} & \textbf{May} & \textbf{Jun} & \textbf{Jul} & \textbf{Aug} & \textbf{Sep} & \textbf{Oct} \\
\hline
\textcolor{Maroon}{5g} & \textcolor{Maroon}{5g} & \textcolor{Maroon}{network} & \textcolor{Maroon}{network} & google & build & \textcolor{Maroon}{5g} & \textcolor{Maroon}{5g} & \textcolor{Maroon}{5g} & \textcolor{Maroon}{5g} & issue & \textcolor{MidnightBlue}{apple} \\
use & \textcolor{Maroon}{network} & \textcolor{Maroon}{5g} & \textcolor{Maroon}{5g} & \textcolor{Maroon}{software} & \textcolor{Maroon}{5g} & \textcolor{Maroon}{us} & \textcolor{Maroon}{us} & \textcolor{Maroon}{network} & \textcolor{Maroon}{india} & \textcolor{MidnightBlue}{oneplus} & use \\
\textcolor{Maroon}{security} & build & \textcolor{Maroon}{equipment} & \textcolor{Maroon}{spy} & \textcolor{MidnightBlue}{apple} & \textcolor{MidnightBlue}{samsung} & \textcolor{Maroon}{security} & \textcolor{Maroon}{security} & \textcolor{Maroon}{countries} & allow & lg & google \\
\textcolor{Maroon}{network} & \textcolor{Maroon}{tech} & \textcolor{Maroon}{national} & \textcolor{Maroon}{us} & devices & use & \textcolor{Maroon}{network} & google & \textcolor{Maroon}{security} & build & \textcolor{MidnightBlue}{pro} & \textcolor{MidnightBlue}{android} \\
\textcolor{MidnightBlue}{products} & \textcolor{MidnightBlue}{brand} & \textcolor{Maroon}{backdoors} & \textcolor{Maroon}{infrastructure} & \textcolor{MidnightBlue}{service} & \textcolor{Maroon}{network} & google & \textcolor{Maroon}{countries} & \textcolor{Maroon}{spy} & \textcolor{Maroon}{risk} & flagship & \textcolor{MidnightBlue}{oneplus} \\
\textcolor{Maroon}{equipment} & \textcolor{MidnightBlue}{samsung} & concern & \textcolor{Maroon}{uk} & \textcolor{MidnightBlue}{products} & google & \textcolor{Maroon}{equipment} & issue & \textcolor{Maroon}{us} & \textcolor{Maroon}{backdoors} & \textcolor{MidnightBlue}{camera} & device \\
\textcolor{Maroon}{tech} & germany & \textcolor{Maroon}{us} & \textcolor{Maroon}{tech} & \textcolor{Maroon}{network} & \textcolor{MidnightBlue}{oneplus} & devices & \textcolor{MidnightBlue}{apple} & \textcolor{Maroon}{national} & \textcolor{Maroon}{military} & forget & \textcolor{Maroon}{equipment} \\
find & \textcolor{MidnightBlue}{oneplus} & allow & \textcolor{Maroon}{equipment} & use & \textcolor{MidnightBlue}{apple} & \textcolor{MidnightBlue}{apple} & \textcolor{Maroon}{tech} & \textcolor{Maroon}{uk} & \textcolor{Maroon}{usa} & \textcolor{MidnightBlue}{mi} & \textcolor{MidnightBlue}{samsung} \\
\textcolor{Maroon}{trust} & find & \textcolor{Maroon}{ban} & \textcolor{Maroon}{military} & support & \textcolor{MidnightBlue}{pro} & work & \textcolor{MidnightBlue}{apps} & \textcolor{Maroon}{equipment} & use & \textcolor{MidnightBlue}{screen} & \textcolor{Maroon}{5g} \\
\textcolor{MidnightBlue}{apple} & \textcolor{MidnightBlue}{apple} & use & \textcolor{Maroon}{security} & \textcolor{MidnightBlue}{brand} & \textcolor{MidnightBlue}{mi} & \textcolor{Maroon}{hardware} & devices & \textcolor{Maroon}{infrastructure} & \textcolor{Maroon}{countries} & \textcolor{Maroon}{5g} & great \\
\textcolor{MidnightBlue}{mate} & devices & \textcolor{Maroon}{country} & intelligence & \textcolor{Maroon}{5g} & \textcolor{MidnightBlue}{brand} & years & \textcolor{Maroon}{equipment} & \textcolor{Maroon}{technology} & \textcolor{Maroon}{western} & devices & \textcolor{MidnightBlue}{apps} \\
\textcolor{Maroon}{steal} & \textcolor{MidnightBlue}{products} & \textcolor{Maroon}{spy} & state & \textcolor{Maroon}{security} & \textcolor{MidnightBlue}{redmi} & \textcolor{Maroon}{canada} & \textcolor{Maroon}{spy} & send & issue & rom & \textcolor{Maroon}{hardware} \\
\textcolor{Maroon}{risk} & \textcolor{Maroon}{ban} & \textcolor{Maroon}{countries} & \textcolor{Maroon}{ban} & \textcolor{MidnightBlue}{pixel} & note & \textcolor{MidnightBlue}{samsung} & \textcolor{Maroon}{canada} & \textcolor{Maroon}{tech} & \textcolor{Maroon}{threaten} & \textcolor{MidnightBlue}{i\textcolor{MidnightBlue}{phone}} & \textcolor{MidnightBlue}{service} \\
\textcolor{Maroon}{national} & \textcolor{Maroon}{equipment} & \textcolor{Maroon}{canada} & \textcolor{Maroon}{countries} & \textcolor{MidnightBlue}{oneplus} & buy & use & allow & \textcolor{Maroon}{threat} & \textcolor{Maroon}{hardware} & worse & \textcolor{MidnightBlue}{camera} \\
devices & \textcolor{Maroon}{security} & devices & \textcolor{Maroon}{government} & \textcolor{Maroon}{technology} & devices & build & \textcolor{Maroon}{backdoors} & \textcolor{MidnightBlue}{apps} & \textcolor{Maroon}{technology} & problem & \textcolor{MidnightBlue}{screen} \\
build & \textcolor{Maroon}{canada} & \textcolor{Maroon}{security} & \textcolor{Maroon}{foreign} & build & \textcolor{Maroon}{equipment} & \textcolor{Maroon}{spy} & \textcolor{Maroon}{steal} & \textcolor{Maroon}{ban} & \textcolor{Maroon}{sanction} & \textcolor{MidnightBlue}{oppo} & instal \\
\textcolor{Maroon}{canada} & \textcolor{Maroon}{technology} & \textcolor{Maroon}{spyware} & \textcolor{Maroon}{national} & \textcolor{MidnightBlue}{screen} & release & \textcolor{Maroon}{trump} & \textcolor{Maroon}{sanction} & \textcolor{Maroon}{steal} & \textcolor{MidnightBlue}{cisco} & great & \textcolor{MidnightBlue}{pixel} \\
\textcolor{Maroon}{access} & evidence & \textcolor{Maroon}{tech} & \textcolor{Maroon}{technology} & \textcolor{Maroon}{tech} & \textcolor{MidnightBlue}{service} & \textcolor{Maroon}{chip} & \textcolor{MidnightBlue}{pixel} & \textcolor{Maroon}{risk} & trade & range & price \\
\textcolor{Maroon}{usa} & \textcolor{Maroon}{steal} & \textcolor{Maroon}{threat} & issue & \textcolor{Maroon}{hardware} & first & \textcolor{Maroon}{uk} & build & \textcolor{Maroon}{arrest} & \textcolor{Maroon}{threat} & terrible & \textcolor{Maroon}{infrastructure} \\
root & \textcolor{MidnightBlue}{redmi} & surveillance & world & cheap & stock & \textcolor{MidnightBlue}{brand} & \textcolor{Maroon}{infrastructure} & users & \textcolor{Maroon}{country} & device & compare \\
\textcolor{Maroon}{infrastructure} & \textcolor{Maroon}{uk} & evidence & allow & intelligence & \textcolor{Maroon}{infrastructure} & \textcolor{MidnightBlue}{mi} & work & user & \textcolor{Maroon}{network} & \textcolor{MidnightBlue}{p20} & \textcolor{MidnightBlue}{miui} \\
concern & lenovo & \textcolor{Maroon}{infrastructure} & \textcolor{Maroon}{threat} & allow & \textcolor{MidnightBlue}{oppo} & \textcolor{MidnightBlue}{oneplus} & \textcolor{Maroon}{national} & concern & \textcolor{Maroon}{international} & \textcolor{Maroon}{hardware} & \textcolor{MidnightBlue}{battery} \\
google & issue & \textcolor{Maroon}{risk} & trade & years & device & device & \textcolor{MidnightBlue}{p40} & collect & political & moto & \textcolor{MidnightBlue}{os} \\
\textcolor{Maroon}{us} & least & build & telecom & avoid & issue & \textcolor{Maroon}{countries} & \textcolor{Maroon}{uk} & \textcolor{Maroon}{india} & destroy & \textcolor{Maroon}{chip} & \textcolor{MidnightBlue}{pro} \\
\textcolor{MidnightBlue}{i\textcolor{MidnightBlue}{phone}} & ericsson & \textcolor{Maroon}{trump} & concern & \textcolor{Maroon}{equipment} & \textcolor{Maroon}{uk} & \textcolor{Maroon}{arrest} & citizens & \textcolor{Maroon}{sanction} & \textcolor{MidnightBlue}{service} & \textcolor{MidnightBlue}{pixel} & annoy\\
\hline
\hline
\multicolumn{2}{|c|}{2020} & \multicolumn{10}{|c|}{2021} \\
\hline
\textbf{Nov} & \textbf{Dec} & \textbf{Jan} & \textbf{Feb} & \textbf{Mar} & \textbf{Apr} & \textbf{May} & \textbf{Jun} & \textbf{Jul} & \textbf{Aug} & \textbf{Sep} & \textbf{Oct} \\
\hline
\textcolor{Maroon}{network} & \textcolor{Maroon}{5g} & \textcolor{Maroon}{5g} & plus & device & \textcolor{Maroon}{australia} & zte & \textcolor{Maroon}{countries} & use & \textcolor{Maroon}{security} & devices & \textcolor{Maroon}{ban} \\
\textcolor{Maroon}{equipment} & \textcolor{Maroon}{network} & \textcolor{Maroon}{hardware} & steam & massive & \textcolor{Maroon}{telecom} & \textcolor{MidnightBlue}{os} & \textcolor{Maroon}{5g} & work & tablet & use & \textcolor{Maroon}{5g} \\
google & \textcolor{Maroon}{spy} & \textcolor{Maroon}{block} & \textcolor{MidnightBlue}{os} & back & install & devices & \textcolor{Maroon}{equipment} & \textcolor{MidnightBlue}{apps} & normal & \textcolor{MidnightBlue}{apple} & \textcolor{Maroon}{security} \\
build & \textcolor{Maroon}{technology} & \textcolor{Maroon}{network} & day & support & \textcolor{Maroon}{network} & \textcolor{Maroon}{hardware} & \textcolor{Maroon}{security} & case & disappoint & \textcolor{MidnightBlue}{i\textcolor{MidnightBlue}{phone}} & \textcolor{Maroon}{us} \\
\textcolor{Maroon}{canada} & \textcolor{Maroon}{tech} & devices & code & \textcolor{MidnightBlue}{phone} & home & \textcolor{MidnightBlue}{pixel} & example & \textcolor{MidnightBlue}{smart\textcolor{MidnightBlue}{phone}} & crash & \textcolor{MidnightBlue}{service} & \textcolor{MidnightBlue}{products} \\
\textcolor{Maroon}{tech} & \textcolor{Maroon}{equipment} & \textcolor{Maroon}{uk} & replace & devices & \textcolor{Maroon}{eavesdrop} & \textcolor{MidnightBlue}{i\textcolor{MidnightBlue}{phone}} & major & avoid & average & \textcolor{MidnightBlue}{pixel} & big \\
sony & intelligence & device & poor & \textcolor{Maroon}{usa} & lock & eu & \textcolor{MidnightBlue}{wifi} & honor & anymore & \textcolor{MidnightBlue}{camera} & \textcolor{Maroon}{equipment} \\
\textcolor{Maroon}{infrastructure} & \textcolor{Maroon}{backdoor} & \textcolor{MidnightBlue}{samsung} & intellectual & important & \textcolor{MidnightBlue}{apps} & work & \textcolor{Maroon}{network} & \textcolor{Maroon}{sanction} & \textcolor{Maroon}{hardware} & \textcolor{MidnightBlue}{mi} & use \\
\textcolor{Maroon}{spy} & agencies & \textcolor{MidnightBlue}{apple} & \textcolor{Maroon}{5g} & info & \textcolor{Maroon}{infrastructure} & \textcolor{MidnightBlue}{ios} & illegal & \textcolor{MidnightBlue}{p30} & past & \textcolor{Maroon}{security} & \textcolor{Maroon}{network} \\
\textcolor{Maroon}{security} & \textcolor{MidnightBlue}{service} & google & life & total & connect & \textcolor{MidnightBlue}{lite} & allow & \textcolor{Maroon}{equipment} & \textcolor{MidnightBlue}{i\textcolor{MidnightBlue}{phone}} & great & \textcolor{Maroon}{steal} \\
interest & \textcolor{Maroon}{national} & watch & assume & \textcolor{Maroon}{america} & fraud & tower & \textcolor{MidnightBlue}{mate} & switch & days & poco & place \\
\textcolor{MidnightBlue}{os} & devices & europe & work & concern & compare & fingerprint & \textcolor{Maroon}{military} & \textcolor{Maroon}{5g} & \textcolor{Maroon}{5g} & device & google \\
zte & request & consider & past & news & warranty & \textcolor{MidnightBlue}{screen} & connect & glad & disable & \textcolor{MidnightBlue}{oneplus} & \textcolor{Maroon}{tech} \\
currently & find & \textcolor{MidnightBlue}{apps} & honest & extreme & tv & \textcolor{Maroon}{software} & build & \textcolor{Maroon}{countries} & instal & \textcolor{MidnightBlue}{brand} & \textcolor{Maroon}{national} \\
whether & \textcolor{Maroon}{chip} & \textcolor{MidnightBlue}{pro} & \textcolor{MidnightBlue}{screen} & supply & none & \textcolor{MidnightBlue}{mate} & \textcolor{Maroon}{trump} & \textcolor{Maroon}{nortel} & cloud & suck & \textcolor{Maroon}{chip} \\
\textcolor{Maroon}{5g} & chinas & lead & billion & \textcolor{Maroon}{network} & attempt & \textcolor{Maroon}{5g} & hack & wrong & provide & \textcolor{MidnightBlue}{oppo} & data \\
create & ask & \textcolor{MidnightBlue}{miui} & \textcolor{Maroon}{uk} & year & \textcolor{Maroon}{5g} & try & \textcolor{Maroon}{technology} & zte & \textcolor{MidnightBlue}{mate} & \textcolor{Maroon}{hardware} & build \\
recommend & claim & \textcolor{MidnightBlue}{wifi} & source & \textcolor{Maroon}{canada} & \textcolor{MidnightBlue}{apple} & \textcolor{MidnightBlue}{p30} & \textcolor{Maroon}{australia} & \textcolor{Maroon}{military} & ads & \textcolor{MidnightBlue}{battery} & \textcolor{Maroon}{trust} \\
\textcolor{Maroon}{military} & \textcolor{Maroon}{countries} & \textcolor{MidnightBlue}{brand} & nearly & days & wait & drop & old & non & \textcolor{MidnightBlue}{p30} & \textcolor{MidnightBlue}{apps} & \textcolor{Maroon}{spy} \\
vivo & back & issue & rip & dollars & devices & push & future & either & totally & end & business \\
\textcolor{Maroon}{sanction} & \textcolor{Maroon}{us} & new & \textcolor{MidnightBlue}{lite} & remember & story & issue & \textcolor{Maroon}{foreign} & personally & components & \textcolor{MidnightBlue}{os} & \textcolor{Maroon}{canada} \\
higher & \textcolor{Maroon}{access} & \textcolor{MidnightBlue}{i\textcolor{MidnightBlue}{phone}} & pick & terrible & different & \textcolor{Maroon}{network} & bite & support & \textcolor{MidnightBlue}{laptop} & experience & product \\
available & \textcolor{Maroon}{western} & budget & question & \textcolor{Maroon}{americans} & \textcolor{MidnightBlue}{service} & \textcolor{Maroon}{chip} & near & due & thank & old & stop \\
use & \textcolor{Maroon}{americans} & ill & face & invest & \textcolor{Maroon}{block} & \textcolor{Maroon}{backdoors} & mine & \textcolor{MidnightBlue}{apple} & wife & \textcolor{MidnightBlue}{products} & trade \\
\textcolor{Maroon}{nortel} & \textcolor{Maroon}{infrastructure} & \textcolor{MidnightBlue}{os} & non & domestic & examples & gb & name & \textcolor{MidnightBlue}{brand} & makers & better & ago\\
\hline
\end{tabular}
}
\label{tab:hauwei-n}
\end{table*}

\begin{table}[h]
\centering
\caption{Top 20 most frequent semantically similar words to ``Huawei'' among negative comments. ``Count" refers to the number of the corresponding word's occurrence in the monthly top 25 most semantically similar words to ``Huawei".}
\begin{tabular}{|c|c|c|c|}
\hline
\textbf{word} & \textbf{count} & \textbf{word} & \textbf{count} \\
\hline
5g & 22 & countries & 9 \\
network & 17 & google & 9 \\
equipment & 15 & hardware & 9 \\
devices & 14 & uk & 8 \\
security & 13 & issue & 8 \\
use & 11 & us & 8 \\
apple & 11 & canada & 8 \\
build & 11 & spy & 8 \\
infrastructure & 10 & technology & 7 \\
tech & 10 & national & 7\\
\hline
\end{tabular}
\label{tab:huawei-n-count}
\end{table}

First, we look into the overall use of words in negative comments about Huawei. Table~\ref{tab:hauwei-n} illustrates the top 25 most similar words to “Huawei” each month from November 1, 2019 to November 1, 2021. Table~\ref{tab:huawei-n-count} shows the top 20 most frequent words in Table~\ref{tab:hauwei-n}. We notice that there are three dominant groups of frequent words in negative comments:
\begin{itemize}
    \item Words closely related to telecommunication technology: “5g”, “network”, “equipment”, “infrastructure”, “tech”, “technology”, “hardware”.
    \item Words related to political entity: “countries”, “uk”, “us”, “canada”.
    \item Words that express concern on network security: “security”, “spy”, “backdoor”, “concern”, “risk”.
\end{itemize}

The frequent occurrence of these three topics as the top most similar words indicates that these words have the most similar context to ``Huawei'' the majority of the time. This leads us to suspect the discussion of Huawei in negative comments may be closely knitted with Huawei's espionage controversy over the allegations of using its telecommunication device to spy on its users for the Chinese government.\footnote{\url{https://www.cfr.org/backgrounder/huawei-chinas-controversial-tech-giant} [Accessed Jan. 8, 2021]} After reviewing comments with these terms as keywords, we find evidence of this suspicion. Many commenters expressed their belief that Huawei is culpable in the spying allegations: 
\begin{itemize}
    \item \textit{“Duh, Huawei IS spying on users, its not rumors, its fact, they have confirmed it, they have whole departments connected to the Chinese military. All Chinese companies are directly or indirectly connected to the Chinese government, telecom ones are of huge interest”}
    \item \textit{“aNd tHe cHiNeSe hUaWeI aRe sPyInG oN uS !!!”}
\end{itemize}

We also observe the word ``ban'' sometimes co-occurs with the three categories above in several months. In examining closer, we discovered that many comments also echo many western governments' ban on Huawei.\footnote{\url{https://www.cnet.com/tech/services-and-software/huawei-ban-timeline-detained-cfo-makes-deal-with-us-justice-department/} [Accessed Jan. 8, 2021]} These comments often characterize Huawei as ``national threat'', ``security risk'', and ``Chinese spyware'', or believed that Huawei has planted ``backdoors'' in their devices. These negative characterizations of Huawei are reflected as the top most similar words of ``Huawei'' in Table~\ref{tab:hauwei-n}. These comments often showed support for a specific country's existing ban on Huawei, or advocate harder sanctions and restrictions on Huawei in the concern and anger of Huawei's potential spying activity. Several example comments are shown in the following:

\begin{itemize}
    \item \textit{“Banning Huawei is normal for any country who cares about national security.”}
    \item \textit{“Given enough time, even Trump can do some good with his malice...eventually...by accident. Now don't forget to ban the rest of the Chinese spyware, ban Huawei from 5g networks, and go after the phones they make.”}
    \item \textit{“YOU PEOPLE need to support banning Huawei in Canada the land of milk and honey!”}
\end{itemize}

While Huawei's telecommunication infrastructure and the alleged spying activity is the dominant theme in the discussion of Huawei throughout our study period, we observe a group of smartphone-related words, such as ``flagship'', ``pro'' (Huawei's smartphone's premium production line), ``camera'', ``screen'' and ``rom'' (Read-Only-Memory), that overtakes telecommunication and spying topics as the top similar words of ``Huawei''. We find that these words are used in the discussion of Huawei's smartphone products, criticizing and showing discontent with Huawei's smartphone products. These comments often complained about certain aspects of smartphones or criticized the company Huawei as a smartphone manufacturer. For instance, a user comments as follows:

\begin{itemize}
    \item \textit{“Huawei Phones, I've had multiple of them and I've experienced nothing but problems. Broken headphone jack, broken charger port, default features crashing the phone, to name a few.”}
\end{itemize}

We analyze the use of words among positive comments using the same methodology, but here we omit the monthly top similar words to ``Huawei'' and only showcase the summarized most frequent top similar words to ``Huawei'' in Table~\ref{tab:huawei-p-count}, as the top similar words are mostly related to only one topic: Huawei's consumer electronics. Table~\ref{tab:huawei-p-count} showcases the top 20 most frequent top similar words to ``Huawei'' from November 1, 2019 to November 1, 2021.

\begin{table}[h]
\centering
\caption{Top 20 most frequent semantically similar words to ``Huawei'' among positive comments. ``Count" refers to the number of the corresponding word's occurrence in the monthly top 25 most semantically similar words to ``Huawei".}
\begin{tabular}{|c|c|c|c|}
\hline
\textbf{word} & \textbf{count} & \textbf{word} & \textbf{count} \\
\hline
p20 & 10 & p40 & 7 \\
mate & 8 & kind & 6 \\
series & 7 & compare & 6 \\
model & 6 & options & 5 \\
service & 5 & change & 5 \\
lite & 5 & problem & 5 \\
least & 5 & believe & 5 \\
move & 5 & nova & 5 \\
love & 5 & cameras & 5 \\
buy & 5 & plus & 5\\
\hline
\end{tabular}
\label{tab:huawei-p-count}
\end{table}

\textit{Consistent} with our finding in the previous topic modeling section, positive comments are mostly associated with consumer electronics, such as smartphones and laptops. As shown in Table~\ref{tab:huawei-p-count}, the most frequent similar words to “Huawei” are either the names of their smartphone models, such as “p20” (Huawei’s 2018 flagship model), “mate” (Huawei’s high-end smartphone series), and “lite” (economic version of the flagship models) or the commonly used words for compliments, such as “love” and “buy”. The comments usually applaud certain aspects of a particular model, and possibly are being posted to recommend this particular model and phone brand to other people:

\begin{itemize}
    \item \textit{“Huawei mate 20 x the best choice, it's very amazing phone, it combines the fantastic design, the very large battery and the best performance. The rest phones are good, but Huawei is the better than them.”}
\end{itemize}

Overall, we have found that among negative comments, Huawei's involvement in the telecommunication industry and 5G network construction have the closest connection to Huawei in the course of the discussion, during which Huawei is denounced as a Chinese spyware company and a security threat. We find that positive comments, on the other hand, focus solely on Huawei's role as a smartphone manufacturer. This finding shows a meaningful difference in how different opinion groups characterize Huawei and provides insights into the reasons behind different sentiments toward Huawei.

\subsection{Case Study 2: Tencent}
To study the temporal pattern of public sentiment toward Tencent, we train one word2vec model for comments on Tencent of each month and outline the most similar words to “Tencent”. Although in the previous study case of Huawei, we conduct and outline the result for positive and negative sentiment separately, we only include the analysis of negative sentiment for Tencent here, for the number of Tencent-related positive comments is small, accounting for less than 0.4\% of the entire corpus. Table~\ref{tab:tencent-n} illustrates the top 25 most similar words to “Tencent” each month from November 1, 2019 to November 1, 2021. Table~\ref{tab:tencent-n-count} shows the top 20 most frequent words in Table~\ref{tab:tencent-n}.

\begin{table*}[htbp]
\centering
\scriptsize
\caption{Top 25 most semantically similar words to ``Tencent'' obtained from the monthly word2vec models of negative comments from November 1, 2019 to November 1, 2021. Words related to Tencent's economic investment are labeled blue. Words related to Tencent's involvement in the gaming industry are labeled green. Words related to Tencent's tie to the Chinese government are labeled red.}
\scalebox{.9}{
\begin{tabular}{|ll|llllllllll|}
\hline
\multicolumn{2}{|c|}{2019}  & \multicolumn{10}{c|}{2020}                                                                                                                          \\ \hline
\textbf{Nov} & \textbf{Dec} & \textbf{Jan} & \textbf{Feb} & \textbf{Mar} & \textbf{Apr} & \textbf{May} & \textbf{Jun} & \textbf{Jul} & \textbf{Aug} & \textbf{Sep} & \textbf{Oct} \\ \hline
\textcolor{MidnightBlue}{money} & \textcolor{MidnightBlue}{money} & \textcolor{MidnightBlue}{money} & \textcolor{OliveGreen}{epic} & studios & \textcolor{OliveGreen}{riot} & \textcolor{OliveGreen}{epic} & \textcolor{OliveGreen}{game} & \textcolor{OliveGreen}{epic} & \textcolor{OliveGreen}{epic} & \textcolor{OliveGreen}{epic} & \textcolor{OliveGreen}{game} \\
give & give & \textcolor{OliveGreen}{mobile} & \textcolor{OliveGreen}{riot} & \textcolor{MidnightBlue}{money} & \textcolor{OliveGreen}{game} & \textcolor{OliveGreen}{riot} & \textcolor{OliveGreen}{epic} & \textcolor{OliveGreen}{game} & \textcolor{MidnightBlue}{money} & fuck & \textcolor{OliveGreen}{riot} \\
care & \textcolor{MidnightBlue}{reddit} & \textcolor{OliveGreen}{play} & \textcolor{OliveGreen}{game} & \textcolor{OliveGreen}{mobile} & trust & \textcolor{OliveGreen}{game} & \textcolor{OliveGreen}{riot} & \textcolor{OliveGreen}{riot} & \textcolor{OliveGreen}{riot} & \textcolor{OliveGreen}{play} & \textcolor{MidnightBlue}{money} \\
\textcolor{OliveGreen}{game} & spend & \textcolor{OliveGreen}{epic} & \textcolor{MidnightBlue}{money} & timi & \textcolor{MidnightBlue}{company} & \textcolor{MidnightBlue}{reddit} & \textcolor{MidnightBlue}{money} & \textcolor{MidnightBlue}{money} & \textcolor{OliveGreen}{play} & \textcolor{MidnightBlue}{money} & \textcolor{OliveGreen}{ggg} \\
\textcolor{OliveGreen}{riot} & make & \textcolor{OliveGreen}{game} & \textcolor{OliveGreen}{play} & pubg & give & \textcolor{MidnightBlue}{money} & \textcolor{MidnightBlue}{reddit} & \textcolor{OliveGreen}{mobile} & \textcolor{OliveGreen}{game} & \textcolor{OliveGreen}{game} & \textcolor{OliveGreen}{epic} \\
\textcolor{OliveGreen}{blizzard} & people & \textcolor{OliveGreen}{skin} & \textcolor{OliveGreen}{blizzard} & \textcolor{OliveGreen}{play} & \textcolor{MidnightBlue}{money} & give & \textcolor{OliveGreen}{play} & stop & fuck & \textcolor{OliveGreen}{mobile} & \textcolor{MidnightBlue}{company} \\
make & fuck & want & \textcolor{OliveGreen}{steam} & \textcolor{OliveGreen}{game} & control & fuck & make & \textcolor{OliveGreen}{play} & \textcolor{MidnightBlue}{reddit} & \textcolor{OliveGreen}{riot} & \textcolor{OliveGreen}{league} \\
free & want & make & spend & spend & \textcolor{MidnightBlue}{reddit} & \textcolor{OliveGreen}{pc} & fuck & fuck & make & \textcolor{MidnightBlue}{reddit} & \textcolor{OliveGreen}{players} \\
way & free & free & \textcolor{OliveGreen}{activision} & give & \textcolor{OliveGreen}{pc} & control & spend & care & \textcolor{OliveGreen}{cod}m & big & pay \\
\textcolor{OliveGreen}{epic} & stop & \textcolor{OliveGreen}{riot} & \textcolor{OliveGreen}{mobile} & \textcolor{OliveGreen}{cheat} & spend & \textcolor{MidnightBlue}{company} & \textcolor{OliveGreen}{cheat} & pubg & \textcolor{OliveGreen}{activision} & hate & make \\
people & partially & \textcolor{OliveGreen}{pc} & free & \textcolor{OliveGreen}{players} & \textcolor{Maroon}{ccp} & shit & \textcolor{OliveGreen}{cheaters} & shit & spend & stake & \textcolor{MidnightBlue}{reddit} \\
fuck & shit & spend & \textcolor{OliveGreen}{devs} & care & \textcolor{OliveGreen}{valorant} & share & people & pay & \textcolor{OliveGreen}{mobile} & give & spend \\
\textcolor{OliveGreen}{play} & \textcolor{MidnightBlue}{invest} & \textcolor{OliveGreen}{players} & care & people & fuck & \textcolor{Maroon}{chinese} & hate & free & shit & \textcolor{OliveGreen}{ggg} & free \\
\textcolor{MidnightBlue}{reddit} & care & fuck & \textcolor{OliveGreen}{players} & guess & \textcolor{Maroon}{government} & \textcolor{MidnightBlue}{invest} & \textcolor{OliveGreen}{activision} & spend & give & make & stop \\
every & stake & give & store & create & american & anti & \textcolor{OliveGreen}{mobile} & give & trust & share & \textcolor{OliveGreen}{play} \\
mean & \textcolor{MidnightBlue}{fund} & \textcolor{MidnightBlue}{reddit} & \textcolor{MidnightBlue}{greedy} & \textcolor{OliveGreen}{hackers} & access & ByteDance & \textcolor{OliveGreen}{players} & pubgm & pubg & spend & \textcolor{OliveGreen}{cheaters} \\
\textcolor{OliveGreen}{ggg} & reason & community & know & doubt & \textcolor{OliveGreen}{play} & \textcolor{OliveGreen}{mobile} & pubg & support & pubgm & pubg & care \\
keep & pay & pubg & \textcolor{OliveGreen}{pc} & scam & \textcolor{Maroon}{chinese} & \textcolor{Maroon}{ccp} & free & \textcolor{OliveGreen}{players} & right & \textcolor{OliveGreen}{cheaters} & \textcolor{OliveGreen}{skin} \\
stop & amount & \textcolor{OliveGreen}{industry} & shit & purpose & know & free & \textcolor{MidnightBlue}{invest} & buy & \textcolor{OliveGreen}{players} & \textcolor{OliveGreen}{hackers} & hand \\
right & \textcolor{OliveGreen}{mobile} & \textcolor{OliveGreen}{activision} & \textcolor{MidnightBlue}{profit} & \textcolor{OliveGreen}{skin} & \textcolor{OliveGreen}{epic} & care & give & \textcolor{OliveGreen}{cheaters} & part & \textcolor{MidnightBlue}{invest} & majority \\
want & \textcolor{OliveGreen}{blizzard} & way & take & \textcolor{OliveGreen}{epic} & support & \textcolor{OliveGreen}{cheat} & \textcolor{MidnightBlue}{sell} & make & \textcolor{OliveGreen}{league} & \textcolor{MidnightBlue}{sell} & ruin \\
\textcolor{MidnightBlue}{invest} & pubg & \textcolor{OliveGreen}{hackers} & \textcolor{OliveGreen}{cheaters} & \textcolor{OliveGreen}{cod} & computer & access & part & nothing & \textcolor{OliveGreen}{skin} & \textcolor{OliveGreen}{cheat} & \textcolor{MidnightBlue}{sell} \\
\textcolor{OliveGreen}{players} & \textcolor{MidnightBlue}{investment} & \textcolor{MidnightBlue}{invest} & try & fuck & data & run & stake & \textcolor{OliveGreen}{hackers} & free & \textcolor{OliveGreen}{activision} & everything \\
say & \textcolor{OliveGreen}{players} & video & make & stop & \textcolor{OliveGreen}{blizzard} & big & \textcolor{OliveGreen}{blizzard} & \textcolor{MidnightBlue}{reddit} & pay & \textcolor{OliveGreen}{steam} & \textcolor{Maroon}{china} \\
boycott & large & try & number & report & rootkit & people & hand & loud & \textcolor{OliveGreen}{steam} & \textcolor{OliveGreen}{players} & blame\\
\hline

\hline
\multicolumn{2}{|c|}{2020}  & \multicolumn{10}{c|}{2021}                                                                                                                          \\ \hline
\textbf{Nov} & \textbf{Dec} & \textbf{Jan} & \textbf{Feb} & \textbf{Mar} & \textbf{Apr} & \textbf{May} & \textbf{Jun} & \textbf{Jul} & \textbf{Aug} & \textbf{Sep} & \textbf{Oct} \\ \hline
\textcolor{OliveGreen}{game} & \textcolor{OliveGreen}{epic} & \textcolor{MidnightBlue}{reddit} & care & give & \textcolor{Maroon}{ccp} & \textcolor{OliveGreen}{riot} & \textcolor{OliveGreen}{epic} & \textcolor{OliveGreen}{riot} & \textcolor{MidnightBlue}{company} & \textcolor{Maroon}{ccp} & \textcolor{MidnightBlue}{money} \\
\textcolor{OliveGreen}{riot} & \textcolor{OliveGreen}{game} & \textcolor{OliveGreen}{riot} & \textcolor{Maroon}{chinese} & \textcolor{OliveGreen}{epic} & \textcolor{OliveGreen}{riot} & \textcolor{MidnightBlue}{company} & \textcolor{MidnightBlue}{reddit} & \textcolor{OliveGreen}{game} & \textcolor{OliveGreen}{game} & \textcolor{OliveGreen}{epic} & fuck \\
\textcolor{OliveGreen}{play} & \textcolor{MidnightBlue}{reddit} & post & \textcolor{MidnightBlue}{reddit} & large & \textcolor{OliveGreen}{game} & \textcolor{OliveGreen}{epic} & \textcolor{Maroon}{ccp} & fuck & \textcolor{Maroon}{chinese} & \textcolor{OliveGreen}{game} & \textcolor{OliveGreen}{play} \\
\textcolor{MidnightBlue}{reddit} & \textcolor{OliveGreen}{riot} & \textcolor{OliveGreen}{game} & fuck & \textcolor{MidnightBlue}{company} & \textcolor{MidnightBlue}{company} & \textcolor{Maroon}{ccp} & people & \textcolor{MidnightBlue}{money} & \textcolor{OliveGreen}{epic} & \textcolor{MidnightBlue}{company} & want \\
\textcolor{MidnightBlue}{money} & \textcolor{MidnightBlue}{money} & fuck & \textcolor{Maroon}{ccp} & \textcolor{MidnightBlue}{money} & \textcolor{MidnightBlue}{reddit} & \textcolor{OliveGreen}{game} & \textcolor{Maroon}{china} & \textcolor{OliveGreen}{epic} & \textcolor{Maroon}{china} & \textcolor{MidnightBlue}{money} & \textcolor{OliveGreen}{game} \\
\textcolor{OliveGreen}{ea} & \textcolor{OliveGreen}{play} & \textcolor{OliveGreen}{activision} & become & stake & \textcolor{OliveGreen}{epic} & fuck & \textcolor{MidnightBlue}{company} & people & \textcolor{Maroon}{ccp} & \textcolor{MidnightBlue}{reddit} & people \\
\textcolor{MidnightBlue}{fund} & \textcolor{OliveGreen}{mihoyo} & \textcolor{OliveGreen}{epic} & \textcolor{OliveGreen}{epic} & force & control & \textcolor{Maroon}{chinese} & \textcolor{OliveGreen}{riot} & make & \textcolor{Maroon}{government} & \textcolor{OliveGreen}{riot} & care \\
\textcolor{OliveGreen}{mobile} & fuck & part & know & product & make & \textcolor{MidnightBlue}{money} & control & \textcolor{MidnightBlue}{company} & \textcolor{MidnightBlue}{money} & hand & keep \\
\textcolor{MidnightBlue}{corporation} & hate & stake & people & share & \textcolor{MidnightBlue}{money} & mean & \textcolor{OliveGreen}{game} & \textcolor{Maroon}{ccp} & \textcolor{OliveGreen}{riot} & \textcolor{Maroon}{china} & give \\
\textcolor{OliveGreen}{league} & \textcolor{MidnightBlue}{invest} & \textcolor{OliveGreen}{league} & \textcolor{OliveGreen}{game} & hard & fuck & \textcolor{MidnightBlue}{reddit} & \textcolor{Maroon}{tie} & \textcolor{OliveGreen}{play} & control & control & everything \\
\textcolor{OliveGreen}{devs} & give & everything & \textcolor{Maroon}{china} & long & \textcolor{Maroon}{china} & \textcolor{Maroon}{china} & ByteDance & \textcolor{Maroon}{china} & business & video & \textcolor{OliveGreen}{epic} \\
\textcolor{Maroon}{tie} & care & small & arm & \textcolor{OliveGreen}{riot} & people & control & \textcolor{MidnightBlue}{invest} & \textcolor{Maroon}{chinese} & \textcolor{Maroon}{party} & \textcolor{Maroon}{tie} & way \\
\textcolor{OliveGreen}{cheat} & \textcolor{OliveGreen}{blizzard} & take & \textcolor{OliveGreen}{fair} & \textcolor{OliveGreen}{play} & involve & \textcolor{Maroon}{government} & give & hand & us & people & know \\
fuck & \textcolor{OliveGreen}{steam} & believe & developers & local & \textcolor{MidnightBlue}{invest} & give & \textcolor{MidnightBlue}{money} & give & right & \textcolor{OliveGreen}{mihoyo} & pay \\
\textcolor{MidnightBlue}{greedy} & \textcolor{OliveGreen}{activision} & \textcolor{MidnightBlue}{money} & bad & \textcolor{MidnightBlue}{fund} & stake & know & \textcolor{MidnightBlue}{fund} & say & \textcolor{OliveGreen}{play} & ByteDance & try \\
\textcolor{OliveGreen}{players} & see & evil & \textcolor{OliveGreen}{league} & \textcolor{OliveGreen}{game} & \textcolor{OliveGreen}{ggg} & take & influence & \textcolor{MidnightBlue}{reddit} & state & give & bad \\
fortnite & part & target & blow & crappy & allow & people & stake & \textcolor{OliveGreen}{league} & \textcolor{Maroon}{communist} & anti & go \\
\textcolor{Maroon}{ccp} & people & right & free & \textcolor{MidnightBlue}{invest} & large & \textcolor{OliveGreen}{league} & reason & big & mean & world & \textcolor{MidnightBlue}{sell} \\
\textcolor{MidnightBlue}{market} & \textcolor{OliveGreen}{ea} & \textcolor{OliveGreen}{industry} & hand & top & care & \textcolor{MidnightBlue}{invest} & free & know & media & take & every \\
partially & shit & leave & \textcolor{MidnightBlue}{money} & decent & entire & care & hand & control & people & \textcolor{MidnightBlue}{invest} & back \\
give & make & alibaba & million & make & \textcolor{OliveGreen}{pc} & big & \textcolor{MidnightBlue}{profit} & stop & \textcolor{MidnightBlue}{reddit} & big & free \\
right & stop & community & see & old & \textcolor{MidnightBlue}{market} & make & \textcolor{MidnightBlue}{investment} & \textcolor{MidnightBlue}{fund} & everything & influence & stock \\
involve & \textcolor{Maroon}{china} & lol & censorship & group & \textcolor{OliveGreen}{mobile} & hand & \textcolor{MidnightBlue}{corporations} & \textcolor{OliveGreen}{mobile} & tech & \textcolor{MidnightBlue}{profit} & time \\
milk & long & \textcolor{MidnightBlue}{greedy} & everyone & biggest & give & hate & share & every & fuck & include & \textcolor{MidnightBlue}{market} \\
\textcolor{OliveGreen}{blizzard} & everything & \textcolor{OliveGreen}{ggg} & give & consider & \textcolor{MidnightBlue}{profit} & \textcolor{MidnightBlue}{market} & data & data & part & hold & shit\\
\hline
\end{tabular}
}
\label{tab:tencent-n}
\end{table*}

\begin{table}[h]
\caption{Top 20 most frequent semantically similar words to ``Tencent" among negative comments. ``Count" refers to the number of the corresponding word's occurrence in the monthly top 25 most semantically similar words to ``Tencent".}
\label{tab:tencent-n-count}
\centering
\begin{tabular}{|c|c|c|c|}
\hline
\textbf{word} & \textbf{count} & \textbf{word} & \textbf{count} \\
\hline
money & 24 & invest & 12 \\
game & 23 & free & 12 \\
epic & 22 & care & 12 \\
riot & 20 & mobile & 12 \\
give & 20 & players & 11 \\
reddit & 20 & company & 10 \\
fuck & 19 & spend & 10 \\
play & 16 & ccp & 10 \\
make & 14 & china & 9 \\
people & 14 & control & 8\\
\hline
\end{tabular}
\end{table}

There are several interesting observations here. First, “money” and “invest” are the underlying theme in the discussion of Tencent during the entire study period, as “money” occurs in every month as the top similar words, and “invest” occurs 50\% of the time. By examining comments with these keywords, we have found they are mostly related to two topics. The first topic focuses on commenters’ discontent toward Tencent’s economic investment, because of Tencent’s Chinese root and Tencent’s alleged close relations with the Chinese government. A portion of the commenters advocate for an economic boycott of Tencent and the companies Tencent has invested in:
\begin{itemize}
    \item  \textit{“What I can do is stay up to date with events here. Delete all blizzard games. Stop buying new blizzard games.  stop playing games that take in Chinese money. Stop watching movies backed by Tencent and other Chinese companies that invest heavily into American entertainment. And continue to support the Hong Kong people here. That's all I can do from the confines of my home. If others are able to go to Hong Kong and help, then they should absolutely do so. They need all the help they can get.”}
    \item \textit{“Well tencent is one of the production companies pumping dirty money into this movie. Maybe china could foot the bill on those tickets instead. YOU WANT TO MAKE A BETTER GLOBAL NEIGHBORHOOD? DONT PAY A CENT TO TENCENT.”}
\end{itemize}

The second topic concerning “money” and “investment” represents commenters’ discontent toward Tencent’s and its invested companies’ monetization behavior in the gaming industry. A group of game-related terms, such as “game”, “riot” (a subsidiary video game company of Tencent), “epic” (a video game company of which 40\% shares are owned by Tencent), and “ggg” (a video game company acquired by Tencent), persist in the top most similar words to ``Tencent'' during the entire study period, and often co-occur with “money” and “investment”. After examining comments with these terms, we have found that commenters usually show discontent toward Tencent and its invested companies’ predatory monetization schemes and disregard for users' gameplay experience. The following example comment showcases a typical comment on this topic:
\begin{itemize}
    \item \textit{“Epic probably wouldn't get such a bad rap for it though, if it wasn't for the fact that most of their monetization systems are predatory and their company is held about 50\% by Tencent, a company also known for predatory monetization (they also designed a great deal of Epic's monetization systems today) and close ties with the Chinese government.”}
\end{itemize}

Another interesting term that persists through the entire study period, occurring as the top similar words to “Tencent” 80\% of the time, is “Reddit”. We have found that there exists large dissatisfaction within the Reddit community toward Tencent’s investment in Reddit itself in February 2019. Many commenters show concerns and discontent about potential censorship, data security issues, and the emergence of communist propaganda on Reddit with Tencent. They express a negative attitude toward Tencent being a shareholder of Reddit because, in their opinion, Tencent is a subordinate company of the Chinese government. The following example comments showcase the typical comments regarding this issue:
\begin{itemize}
    \item \textit{“Reddit is owned by the CCP firm Tencent. China didn't take any vaccines. This is a Propaganda containment site.”}
    \item \textit{“China partially own reddit through tencent, there's a squadron of power users who just mindlessly defend the CCP and Chinese interests across the site.”}
\end{itemize}
 
Our last observation is through monitoring the change of top similar words to “Tencent”.  As we have described above, the discussion around Tencent’s economic investment and Tencent’s monetization behavior are the underlying themes in the entire discussion, but we can also observe that in certain months, the top similar words composition deviates from these underlying themes. For instance, in April 2020, the top similar words to “Tencent” are a group of words that explicitly refer to China, such as “Chinese”, “CCP”, and “government”, which do not exist in the preceding five months. We have found that a particular news event has triggered a backlash in the Reddit community: Tencent’s fully-owned video game company Riot has released a video game called Valorant, which has been found to have a built-in invasive anti-cheat system, that runs continuously even when the game is not booted.  Many comments express concern and anger toward this invasive anti-cheat system, drawing links to the Riot’s parent company Tencent and accusing Tencent of deliberately planting the anti-cheat system as spyware to collect user information for the Chinese government:
\begin{itemize}
    \item \textit{“··· I wasn’t planning to install Valorant or any other Tencent/CCP China spyware, but good to have confirmation that’s the right decision.”}
    \item \textit{“Giving people freedom? That’s not how Tencent is used to running games because that’s not how Chinese are used to running a society. Remember you’re working with an entity ONE HUNDRED PERCENT owned by Tencent, a Chinese state-run mass media corporation.”}
\end{itemize}

Overall, these findings provide a detailed understanding of how and why the Reddit community expresses negative sentiments toward Tencent. The discontent toward Tencent’s economic investment, its predatory monetization behavior, and its tie to the Chinese government often intertwine to a degree that the commenters rarely express their negative view on Tencent in one single aspect. These findings also show that the reasons behind users’ negative views are very consistent throughout our study period. It is because that Tencent’s economic investment, its predatory monetization behavior, and its tie to the Chinese government are intrinsic to the company itself, which are Tencent’s business model and its Chinese origin, and thus unaffected by local events. Therefore, we anticipate the same pattern of public opinion toward Tencent will persist in the near future because Tencent most likely will  not change its business model or cut ties with the Chinese government.

\subsection{Case Study 3: TikTok}
\begin{table*}[htbp]
\centering
\scriptsize
\caption{Top 25 most semantically similar words to ``TikTok'' obtained from the monthly word2vec models of negative comments from November 1, 2019 to November 1, 2021. Words related to Tiktok's social media platform, user community, and user content are labeled blue. Words related to China, the Chinese government, and the spying controversy are labeled red.}
\scalebox{.9}{
\begin{tabular}{|ll|llllllllll|}
\hline
\multicolumn{2}{|c|}{2019}  & \multicolumn{10}{c|}{2020}                                                                                                                          \\ \hline
\textbf{Nov} & \textbf{Dec} & \textbf{Jan} & \textbf{Feb} & \textbf{Mar} & \textbf{Apr} & \textbf{May} & \textbf{Jun} & \textbf{Jul} & \textbf{Aug} & \textbf{Sep} & \textbf{Oct} \\ \hline
Bytedance & see & \textcolor{MidnightBlue}{people} & \textcolor{MidnightBlue}{hate} & \textcolor{MidnightBlue}{hate} & \textcolor{MidnightBlue}{hate} & \textcolor{MidnightBlue}{hate} & \textcolor{MidnightBlue}{hate} & \textcolor{MidnightBlue}{apps} & \textcolor{Maroon}{ban} & \textcolor{Maroon}{ban} & \textcolor{MidnightBlue}{hate} \\
call & \textcolor{MidnightBlue}{post} & \textcolor{MidnightBlue}{memes} & \textcolor{MidnightBlue}{people} & \textcolor{MidnightBlue}{content} & \textcolor{MidnightBlue}{post} & \textcolor{MidnightBlue}{\textcolor{MidnightBlue}{\textcolor{MidnightBlue}{video}}s} & \textcolor{MidnightBlue}{people} & \textcolor{MidnightBlue}{app} & american & \textcolor{MidnightBlue}{app} & \textcolor{OliveGreen}{instagram} \\
result & censorship & \textcolor{MidnightBlue}{post} & \textcolor{MidnightBlue}{fuck} & \textcolor{OliveGreen}{reddit} & \textcolor{MidnightBlue}{bad} & \textcolor{MidnightBlue}{people} & \textcolor{MidnightBlue}{platform} & information & \textcolor{MidnightBlue}{user} & tiktoks & reason \\
\textcolor{MidnightBlue}{share} & \textcolor{MidnightBlue}{dumb} & funny & reason & \textcolor{MidnightBlue}{post} & \textcolor{MidnightBlue}{\textcolor{MidnightBlue}{\textcolor{MidnightBlue}{video}}s} & \textcolor{MidnightBlue}{bad} & \textcolor{MidnightBlue}{users} & collect & collect & say & \textcolor{MidnightBlue}{social} \\
purchase & hand & \textcolor{MidnightBlue}{content} & \textcolor{MidnightBlue}{app} & \textcolor{MidnightBlue}{people} & fake & \textcolor{MidnightBlue}{content} & \textcolor{MidnightBlue}{app} & \textcolor{OliveGreen}{facebook} & \textcolor{Maroon}{spy} & fact & bc \\
financial & propaganda & \textcolor{MidnightBlue}{shit} & \textcolor{OliveGreen}{reddit} & hat & see & kid & \textcolor{OliveGreen}{instagram} & \textcolor{Maroon}{spyware} & information & wechat & \textcolor{OliveGreen}{facebook} \\
interest & need & \textcolor{MidnightBlue}{\textcolor{MidnightBlue}{\textcolor{MidnightBlue}{video}}s} & \textcolor{MidnightBlue}{bad} & \textcolor{MidnightBlue}{\textcolor{MidnightBlue}{\textcolor{MidnightBlue}{video}}s} & stuff & \textcolor{MidnightBlue}{post} & \textcolor{MidnightBlue}{content} & access & \textcolor{OliveGreen}{facebook} & nation & friends \\
\textcolor{MidnightBlue}{platform} & citizens & \textcolor{MidnightBlue}{cringe} & \textcolor{MidnightBlue}{cringe} & \textcolor{MidnightBlue}{cringe} & please & \textcolor{MidnightBlue}{shit} & hat & \textcolor{MidnightBlue}{users} & \textcolor{Maroon}{steal} & \textcolor{OliveGreen}{facebook} & \textcolor{MidnightBlue}{stupid} \\
\textcolor{MidnightBlue}{content} & power & \textcolor{MidnightBlue}{stupid} & fortnite & see & \textcolor{MidnightBlue}{app} & \textcolor{MidnightBlue}{platform} & \textcolor{MidnightBlue}{dance} & \textcolor{Maroon}{spy} & us & children & toxic \\
foreign & \textcolor{MidnightBlue}{hate} & kid & use & sub & \textcolor{MidnightBlue}{content} & \textcolor{OliveGreen}{instagram} & opinion & american & Bytedance & microsoft & attention \\
mainland & \textcolor{MidnightBlue}{\textcolor{MidnightBlue}{\textcolor{MidnightBlue}{video}}s} & \textcolor{MidnightBlue}{cancer} & \textcolor{MidnightBlue}{\textcolor{MidnightBlue}{video}} & kid & \textcolor{MidnightBlue}{cringe} & \textcolor{MidnightBlue}{cringe} & info & \textcolor{Maroon}{steal} & access & censor & watch \\
gain & \textcolor{MidnightBlue}{content} & fortnite & \textcolor{MidnightBlue}{shit} & censor & \textcolor{MidnightBlue}{community} & \textcolor{MidnightBlue}{dumb} & censor & \textcolor{Maroon}{ban} & \textcolor{Maroon}{china} & \textcolor{MidnightBlue}{people} & say \\
delete & foreign & ruin & \textcolor{MidnightBlue}{awful} & \textcolor{MidnightBlue}{social} & sub & funny & \textcolor{MidnightBlue}{cringe} & \textcolor{Maroon}{beijing} & \textcolor{Maroon}{ccp} & laws & \textcolor{OliveGreen}{twitter} \\
regime & \textcolor{MidnightBlue}{share} & \textcolor{MidnightBlue}{popular} & kid & \textcolor{MidnightBlue}{fuck} & \textcolor{MidnightBlue}{shit} & \textcolor{MidnightBlue}{stupid} & \textcolor{MidnightBlue}{post} & tech & citizens & \textcolor{MidnightBlue}{users} & dislike \\
million & full & fair & \textcolor{MidnightBlue}{platform} & \textcolor{MidnightBlue}{popular} & love & \textcolor{MidnightBlue}{app} & stuff & data & microsoft & \textcolor{MidnightBlue}{user} & wrong \\
harvest & essentially & guy & mean & \textcolor{OliveGreen}{instagram} & everyone & mean & censorship & law & \textcolor{MidnightBlue}{users} & personal & music \\
via & everyone & \textcolor{MidnightBlue}{platform} & \textcolor{MidnightBlue}{post} & censorship & \textcolor{MidnightBlue}{annoy} & fortnite & agree & \textcolor{MidnightBlue}{share} & force & deal & general \\
supporters & \textcolor{MidnightBlue}{memes} & enjoy & deserve & toxic & \textcolor{OliveGreen}{instagram} & censorship & overlords & us & tech & majority & type \\
conglomerate & heavily & sub & everything & delete & \textcolor{MidnightBlue}{comment} & censor & machine & \textcolor{Maroon}{ccp} & law & applause & ass \\
youtube & connect & \textcolor{OliveGreen}{instagram} & good & \textcolor{MidnightBlue}{bad} & \textcolor{OliveGreen}{redditors} & love & \textcolor{MidnightBlue}{bad} & \textcolor{MidnightBlue}{user} & \textcolor{Maroon}{trump} & funny & hurt \\
\textcolor{MidnightBlue}{fuck} & ly & \textcolor{MidnightBlue}{annoy} & simply & \textcolor{MidnightBlue}{app} & show & \textcolor{OliveGreen}{facebook} & view & \textcolor{MidnightBlue}{private} & laws & clear & girls \\
censor & collect & \textcolor{MidnightBlue}{\textcolor{MidnightBlue}{video}} & ruin & Byte\textcolor{MidnightBlue}{dance} & things & concentration & children & send & \textcolor{MidnightBlue}{share} & ask & suck \\
mass & political & close & \textcolor{MidnightBlue}{social} & \textcolor{MidnightBlue}{cancer} & half & \textcolor{MidnightBlue}{community} & kid & \textcolor{MidnightBlue}{social} & wechat & \textcolor{Maroon}{threat} & although \\
obviously & site & \textcolor{MidnightBlue}{community} & \textcolor{MidnightBlue}{\textcolor{MidnightBlue}{video}s} & spread & hell & \textcolor{MidnightBlue}{dance} & promote & \textcolor{Maroon}{privacy} & need & \textcolor{MidnightBlue}{share} & sad \\
collect & send & easy & funny & evil & \textcolor{MidnightBlue}{platform} & delete & see & technology & \textcolor{MidnightBlue}{private} & involve & child\\
\hline

\hline
\multicolumn{2}{|c|}{2020}  & \multicolumn{10}{c|}{2021}                                                                                                                          \\ \hline
\textbf{Nov} & \textbf{Dec} & \textbf{Jan} & \textbf{Feb} & \textbf{Mar} & \textbf{Apr} & \textbf{May} & \textbf{Jun} & \textbf{Jul} & \textbf{Aug} & \textbf{Sep} & \textbf{Oct} \\ \hline
\textcolor{MidnightBlue}{shit} & \textcolor{OliveGreen}{facebook} & hat & news & months & legal & \textcolor{MidnightBlue}{platform} & collect & \textcolor{MidnightBlue}{dumb} & \textcolor{MidnightBlue}{fuck} & public & invest \\
\textcolor{MidnightBlue}{post} & \textcolor{OliveGreen}{twitter} & process & past & full & guy & \textcolor{MidnightBlue}{hate} & harvest & breach & \textcolor{MidnightBlue}{\textcolor{MidnightBlue}{video}} & disappear & \textcolor{MidnightBlue}{private} \\
site & hat & gamble & kind & group & \textcolor{Maroon}{threat} & dislike & intelligence & \textcolor{MidnightBlue}{\textcolor{MidnightBlue}{video}} & \textcolor{MidnightBlue}{people} & \textcolor{MidnightBlue}{\textcolor{MidnightBlue}{platforms}} & servers \\
\textcolor{MidnightBlue}{people} & \textcolor{MidnightBlue}{post} & \textcolor{MidnightBlue}{content} & add & term & grow & \textcolor{MidnightBlue}{social} & internet & \textcolor{MidnightBlue}{stupid} & \textcolor{OliveGreen}{reddit} & \textcolor{MidnightBlue}{hate} & \textcolor{OliveGreen}{facebook} \\
ea & \textcolor{MidnightBlue}{stupid} & emote & include & film & yu & \textcolor{MidnightBlue}{cringe} & matter & face & take & toxic & american \\
guy & \textcolor{MidnightBlue}{people} & learn & show & massive & friends & laws & govt & short & everything & attention & parent \\
real & \textcolor{MidnightBlue}{bad} & mind & kid & build & history & fee & foreign & exactly & call & propaganda & operate \\
\textcolor{MidnightBlue}{\textcolor{MidnightBlue}{\textcolor{MidnightBlue}{video}}s} & kind & \textcolor{MidnightBlue}{\textcolor{MidnightBlue}{\textcolor{MidnightBlue}{video}}s} & \textcolor{MidnightBlue}{post} & million & chance & \textcolor{MidnightBlue}{\textcolor{MidnightBlue}{\textcolor{MidnightBlue}{video}}s} & military & \textcolor{OliveGreen}{facebook} & pretty & internet & act \\
\textcolor{MidnightBlue}{hate} & \textcolor{MidnightBlue}{\textcolor{MidnightBlue}{platforms}} & behind & \textcolor{MidnightBlue}{\textcolor{MidnightBlue}{\textcolor{MidnightBlue}{video}}s} & local & abuse & asian & access & wrong & try & political & directly \\
say & ruin & \textcolor{MidnightBlue}{awful} & ui & lead & send & wrong & allow & anyway & make & believe & foreign \\
person & \textcolor{MidnightBlue}{comment} & improve & criticize & hold & political & whenever & large & law & infrastructure & dude & \textcolor{OliveGreen}{reddit} \\
\textcolor{MidnightBlue}{fuck} & seriously & corrupt & sub & anti & suppose & owners & \textcolor{MidnightBlue}{people} & depend & influence & monitor & listen \\
\textcolor{MidnightBlue}{platform} & \textcolor{OliveGreen}{instagram} & proof & Byte\textcolor{MidnightBlue}{dance} & kind & \textcolor{MidnightBlue}{cringe} & \textcolor{MidnightBlue}{song} & \textcolor{MidnightBlue}{private} & board & hard & sense & clearly \\
hear & \textcolor{MidnightBlue}{hate} & completely & start & hell & entire & god & tie & ones & western & limit & fund \\
\textcolor{MidnightBlue}{annoy} & tik & argue & assume & better & collect & ask & corporations & corporation & start & privilege & piss \\
codm & account & credit & dead & \textcolor{MidnightBlue}{content} & complain & important & citizens & censorship & \textcolor{MidnightBlue}{post} & deserve & businesses \\
million & whole & happy & behind & \textcolor{MidnightBlue}{\textcolor{MidnightBlue}{video}} & counter & fine & investment & subject & interest & western & include \\
pretend & incredibly & sake & edit & different & damn & violations & heavily & authoritarian & jack & negative & position \\
eye & hell & matter & sad & fine & clear & host & request & \textcolor{MidnightBlue}{content} & everyone & pressure & prove \\
face & team & sit & rip & wrong & follow & conspiracy & understand & harvest & stop & \textcolor{MidnightBlue}{comment} & news \\
history & trend & entire & day & pubg & track & society & talk & include & trade & kid & help \\
subreddit & communism & message & produce & riot & server & media & completely & \textcolor{MidnightBlue}{awful} & little & reduction & infiltrate \\
fortnite & whatever & everyone & lie & deny & collection & article & tell & hide & things & damage & \textcolor{Maroon}{privacy} \\
holy & white & later & guy & \textcolor{MidnightBlue}{\textcolor{MidnightBlue}{\textcolor{MidnightBlue}{video}}s} & prove & reality & american & address & real & call & outside \\
\textcolor{MidnightBlue}{comment} & studio & minority & person & \textcolor{MidnightBlue}{users} & tik & regardless & profit & hat & human & \textcolor{OliveGreen}{twitter} & forget\\
\hline
\end{tabular}
}
\label{tab:tiktok-n}
\end{table*}

\begin{table}[h]
\centering
\caption{Top 20 most frequent semantically similar words to ``TikTok'' among negative comments. ``Count" refers to the number of the corresponding word's occurrence in the monthly top 25 most semantically similar words to ``TikTok".}
\label{tab:tiktok-n-count}
\begin{tabular}{|c|c|c|c|}
\hline
\textbf{word} & \textbf{count} & \textbf{word} & \textbf{count} \\
\hline
videos & 11 & instagram & 7 \\
hate & 11 & collect & 6 \\
post & 11 & bad & 6 \\
people & 10 & share & 5 \\
content & 10 & censor & 5 \\
cringe & 8 & users & 5 \\
platform & 8 & social & 5 \\
facebook & 8 & stupid & 5 \\
app & 7 & shit & 5 \\
kid & 7 & censorship & 5\\
\hline
\end{tabular}
\end{table}

Using the same methodology, we study the Reddit users' use of words related to the company ByteDance. Note here although we are interested in studying the company ByteDance itself, we use the word ``Tiktok'', as a majority of people refer to the company as ``Tiktok'' rather than ``ByteDance''. We also omit the analysis of Tiktok-related positive comments, for the quantity of Tiktok-related positive comments is small, with less than 200 comments in total, to conduct any meaningful analysis. Therefore we only outline the top 25 words that are semantically similar to ``Tiktok'' among negative comments in Table~\ref{tab:tiktok-n} and summarize the top 20 most frequently occurred terms in Table~\ref{tab:tiktok-n-count}.

Our attention is first caught by a set of words including “hate”, “videos”, “content”, “cancer”, “platform”, “bad”, “stupid”, “shit”, \textit{etc}, which frequently occur during the entire study period. Given that TikTok is a video-focused social networking platform that hosts a variety of short-form user videos,\footnote{\url{https://en.wikipedia.org/w/index.php?title=TikTok\&oldid=1059570012} [Accessed Dec. 23, 2021]} a connection can be drawn between Reddit users’ negative attitude and the popular videos on TikTok as well as the Tiktok user community. We find that many commenters expressed harsh criticism toward the contents of Tiktok and the whole Tiktok community, referring to both as ``toxic''. For instance, some comments stated:  
\begin{itemize}
    \item \textit{“Wtf is wrong with people on tiktok. I’ve seen some of the most toxic shit on that app”}
    \item \textit{“Can't change the Fact that TikTok is Shit and just stupid”}.
\end{itemize}

From the above, we can see that the discussion of Tiktok is closely surrounding its role as a social media platform, its user community, and the user-generated content on the platform. This is very \textbf{different} from the case of Huawei and Tencent, as the users' characterization of Tiktok has less emphasis on Tiktok's Chinese origin and its ties to the Chinese government. Another evidence of this fact is that Tiktok is discussed in a similar context with other non-Chinese social media platforms, such as ``Twitter", ``Facebook", ``Reddit", and ``Instagram" rather than its Chinese counterpart, such as ``Huawei" and ``Tencent".  We can observe from Table~\ref{tab:tiktok-n} that these non-Chinese social media platforms appeared as Tiktok's most similar words most of the time. We also find evidence of this in many comments, where users put Tiktok and these companies in juxtaposition: 

\begin{itemize}
    \item \textit{“Twitter is great, Reddit too of course. IG is ok but TikTok is mostly cringe”}
    \item \textit{“*[sigh]* I hate Twitter. And Facebook. And Instagram. And Tumblr. And TikTok. Yes, even Reddit sometimes."}
\end{itemize}

While Tiktok is discussed as a social media platform and not characterized closely with China most time during our study periods, we do observe certain exceptional time periods, where Tiktok became closely related to China and the Chinese government. One such period is from July 2020 to September 2020. Not only does “ban” become the most related word with “Tiktok”, other words including “spy”, “ccp”, “china”, and “collect” occur. We believe this change is closely related to Donald Trump's announcement in July 2020 to divest China's ByteDance’s ownership of Tiktok, and that the US government was considering banning TikTok.\footnote{\url{https://www.cnbc.com/2020/07/31/trump-says-he-will-ban-tiktok-through-executive-action-as-soon-as-saturday.html} [Accessed Jan. 8, 2022]} We have found that many Reddit users resonate with this political event. Many users show support for the Tiktok ban. For instance, one commenter notes: \textit{“TikTok is cancer and Trump is right to ban it, change my mind.”}  The most dominant concern is unsurprisingly the same as the concerns for Tencent and Huawei: data security issues and the alleged possibility that Tiktok may be utilized by the Chinese government to spy on its users:
\begin{itemize}
    \item \textit{“tiktok was not banned for being Chinese, it was banned for being a chinese spy tool”}
    \item \textit{“People be like but TikTok was my lifetime career you can’t just ban it, I’d rather lose my job than have fucking China spy on all of my shit”}
    \item \textit{“It's being banned because it's spyware that has been providing user data to the CCP while censoring anything critical of them. That's not a conspiracy, it's what they've said they're doing.  You seriously need to get informed before you speak on this any more.  Tiktok is a terrible company.”}
\end{itemize}

Overall, the negative comments on Tiktok mostly surround the social media platform, platform community, and user-generated content, where many commenters express harsh criticism to these three aspects and describe them as toxic and stupid. We also find for most of the time, Reddit users' characterization of Tiktok has a meaningful difference from the way they characterize Huawei and Tencent, where they mostly refer to Tiktok as a general social media platform and put less emphasis on Tiktok's Chinese origin and ties with the Chinese government. However, real-world political events may completely change this characterization in certain periods, such as when Trump announced his plan to ban Tiktok for national security concerns in July 2020, and on Reddit, Tiktok became closely knitted with China, the Chinese government and a term related to spying activities. These findings show interesting insights into Reddit users' characterization of Tiktok (ByteDance). They show how Tiktok (ByteDance) is viewed differently than other Chinese technology companies in terms of its closeness to China and the Chinese government, and how this difference could be quickly changed by the influence of governmental actions.

\section{Discussion and Conclusion}
In this paper, we present a fine-grained study of public opinion toward Chinese technology companies on Reddit using computational methods. We employ a state-of-the-art transformer model to classify our data into three categories: positive, negative, and irrelevant. We find that the public generally shows more negative attitude than positive attitude toward Chinese technology companies. Next, we use LDA to model the latent topics among negative comments and positive comments, respectively.  Positive comments are usually associated with the companies’ consumer products, such as smartphones and laptops. We also found that negative comments have a more diverse topic distribution. Notable topics include criticism of the platform, dissatisfaction with companies' products, concerns over companies’ ties to the Chinese government, data security and 5G construction, and general political discussions. 

By investigating the most semantically similar words in the comments to the names of three Chinese technology companies over time, we have discovered the predominant themes, and identified several real-world events that affect the public’s characterization of these companies. We take the top 3 companies that have the largest volume of comments as our study cases: Huawei, Tencent, and Tiktok. We study the negative comments and positive comments separately. 

For Huawei, we have found that negative comments are heavily focused on its role as a telecommunication device provider throughout our study period. The most dominant concern/discontent about the company rests on its alleged threat of spying activities. Positive comments, on the other hand, focus on its role as a smartphone manufacturer. The most frequent similar words to ``Huawei'' are the names of its smartphone products and common compliment words for smartphones. 

For Tencent, we have found that negative comments are usually associated with Tencent’s economic investment, predatory monetization behavior, and its tie to the Chinese government. These themes among the negative comments have dominated the online discussion and are consistent during our study period. 

Concerning the public's characterization of Tiktok, we have found that negative views are usually targeted toward this social media platform, the platform community and the content. Furthermore, our analysis reveals that Tiktok can be distinguished by its relatively weaker association with China and the Chinese government compared to other social media platforms overall, with only certain exceptional time periods affected by political events, such as when Trump announced his plan to ban Tiktok in the United States in July 2020. Tiktok became tightly connected to China, the Chinese government, and spying allegations as its fellow Chinese technology companies in the following two months. 

Overall, we have found that Reddit users hold more negative sentiment toward Chinese technology companies. We have characterized users’ concerns and discontent, and explored the reasons behind them. These findings serve to increase our understanding of the wide public opinion toward Chinese technology companies both in the political sense and commercial sense. We believe the findings of our study can be of interest to those who attempt to understand public sentiment toward Chinese technology companies and could provide useful insights into the reasons behind the public sentiment.

Our study has its limitations. For instance, although our XLNet classifier achieves acceptable results in classifying all three categories, we believe the performance of the classifier can still be improved especially in predicting negative and positive comments. In the future, we intend to address this issue by obtaining a larger and more balanced training dataset. Apart from that, another concern is the potential data collection biases that are introduced by the ranking algorithm of Reddit.\footnote{\url{https://github.com/reddit-archive/reddit} [Accessed Aug. 14, 2022]} Reddit's algorithm (1) considers early submissions more important, and (2) favors topics that have an overwhelming opinion (i.e., The topic is less controversial where most users hold similar opinions). To verify if the sentiment distributions vary across different data collection periods, we conduct additional analysis on the sentiment values in different periods - pre-COVID (October to November 2019) and during COVID (2020 to 2021). To account for seasonality, we choose the same two months in 2020 and 2021. We find that the sentiment compositions of Reddit comments are \textit{consistent} across the three periods. There are approximately 1.5 times more negative comments than positive comments. We also intend to conduct our analysis using Twitter data. Although the discussion quality is higher on Reddit, Twitter, as a more popular platform, can allow us to access broader-scale opinions toward our study subjects. 

As the first work to characterize public opinion on Chinese technology companies, our study has the following broader implications. First of all, unlike the previous studies that focus on the relations between Chinese technology companies and the Western governments~\cite{huawei_mwz,hue_cry_huawei}, we shed light on the reasons behind the public sentiment, especially behind the public’s discontent toward these companies in a {\it commercial} sense, such as the discussions around the companies’ consumer products and economic investment. The finding of our study can provide useful information to not only these companies and other general Chinese companies that seek to expand their overseas market, to assess their companies’ image in areas outside China, but also companies that do business with Chinese companies. Second, our study also reveals a wide  mistrust toward Chinese technology companies out of political and ideological concern, and identifies several dominant aspects of such mistrust, such as the data privacy concern and companies’ close ties to the Chinese government. The finding of our studies can be informative to scholars who study China’s foreign relations and China’s emerging technology and economic influence in the global sphere. 

\bibliography{report}

\end{document}